\documentclass[lettersize,journal]{IEEEtran}
\def\BibTeX{{\rm B\kern-.05em{\sc i\kern-.025em b}\kern-.08em
    T\kern-.1667em\lower.7ex\hbox{E}\kern-.125emX}}
\usepackage{cite}
\usepackage{algpseudocode}
\usepackage{graphicx}
\usepackage{amsmath}
\usepackage{amsfonts}
\usepackage{epstopdf}
\usepackage{xcolor}
\usepackage{enumerate}
\usepackage{algorithm}
\usepackage{multirow}
\usepackage{transparent}
\usepackage{subfigure}
\usepackage{float}

\newtheorem{theorem}{Theorem}

\newtheorem{corollary}{Corollary}
\newtheorem{lemma}{Lemma}

\newcommand{\T}{\mathrm{T}}
\newcommand{\R}{\mathrm{R}}

\newcommand{\e}{\textnormal{e}}

\let\ss= \scriptscriptstyle
\begin{document}
	\title{Analysis of MC Systems Employing Receivers Covered by Heterogeneous Receptors}
\author{\thanks{This paper was accepted for presentation in part at the 2022 IEEE International Conference on Communication (ICC) \cite{huang2021analysis}.}Xinyu Huang\thanks{X. Huang and N. Yang are with the School of Engineering, Australian National University, Canberra, ACT 2600, Australia (e-mail: xinyu.huang1@anu.edu.au; nan.yang@anu.edu.au)}, \textit{Student Member, IEEE}, Yuting Fang\thanks{Y. Fang is with the Department of Electrical and Electronic Engineering, S. T. Johnston is with the Systems Biology Laboratory, School of Mathematics and Statistics, and Department of Biomedical Engineering, and M. Faria is with the Department of Biomedical Engineering and ARC Centre of Excellence in Covergent Bio-Nano Science and Technology, The University of Melbourne, Parkville, VIC 2010, Australia (e-mail: yuting.fang@unimelb.edu.au; stuart.johnston@unimelb.edu.au; matthew.faria@unimelb.edu.au)}, \textit{Student Member, IEEE}, Stuart T. Johnston, Matthew Faria, Nan Yang, \textit{Senior Member, IEEE} and Robert Schober\thanks{R. Schober is with the Institute for Digital Communications, Friedrich-Alexander-University Erlangen-N{\"u}rnberg (FAU), 91054 Erlangen, Germany (e-mail: robert.schober@fau.de)}, \textit{Fellow, IEEE}}
\maketitle
\begin{abstract}
	This paper investigates the channel impulse response (CIR), i.e., the molecule hitting rate, of a molecular communication (MC) system employing an absorbing receiver (RX) covered by multiple non-overlapping receptors. In this system, receptors are heterogeneous, i.e., they may have different sizes and arbitrary locations. Furthermore, we consider two types of transmitter (TX), namely a point TX and a membrane fusion (MF)-based spherical TX. We assume the point TX or the center of the MF-based TX has a fixed distance to the center of the RX. Given this fixed distance, the TX can be at different locations and the CIR of the RX depends on the exact location of the TX. By averaging over all possible TX locations, we analyze the expected molecule hitting rate at the RX as a function of the sizes and locations of the receptors, where we assume molecule degradation may occur during the propagation of the signaling molecules. Notably, our analysis is valid for different numbers, a wide range of sizes, and arbitrary locations of the receptors, and its accuracy is confirmed via particle-based simulations. Exploiting our numerical results, we show that the expected number of absorbed molecules at the RX increases with the number of receptors, when the total area on the RX surface covered by receptors is fixed. Based on the derived analytical expressions, we compare different geometric receptor distributions by examining the expected number of absorbed molecules at the RX. We show that evenly distributed receptors result in a larger number of absorbed molecules than other distributions. We further compare three models that combine different types of TXs and RXs. Compared to the ideal model with a point TX and a fully absorbing RX, the practical model with an MF-based TX and an RX with heterogeneous receptors yields a lower peak CIR, suffers from more severe inter-symbol interference, and gives rise to a higher average bit error rate (BER). This underlines the importance of our analysis of practical TX-RX models since the existing CIR and BER analyses based on the ideal model do not reflect the performance achievable in practice.
\end{abstract}

\section{Introduction}
The Internet of Bio-Nano Things (IoBNT) is a new networking paradigm for interconnecting modified biological organisms, e.g., cells and bacteria, which has huge potential for improving human healthcare and water pollution control \cite{akyildiz2015internet}. Inspired by nature, molecular communication (MC) is envisioned to be a promising solution for realizing communication at nanoscale \cite{farsad2016comprehensive}. In MC, molecules are released by a transmitter (TX), and then propagate in a fluid medium until they arrive at a receiver (RX). The RX detects and decodes the information encoded in the molecules. Therefore, accurate modeling of practical TXs and RXs is crucial for the analysis and design of end-to-end MC systems.

In biology, living cells recognize molecules via receptor-ligand interactions, which are fundamental for cells to communicate with their neighbors and the entire organism \cite{guryanov2016receptor}. Specifically, a signal is observed by a cell by binding of ligands to their complementary receptors, which initiates a cascade of chemical reactions. Despite the complexity of the actual reception process, the RX models proposed in the literature are relatively simple. For example, the most widely-adopted RX models in MC are passive RX, fully absorbing RX, and reactive RX \cite{jamali2019channel}. The passive RX ignores receptor-ligand interactions, while the fully absorbing and reactive RXs ignore the receptor size and assume that an infinite number of receptors cover the entire RX surface. To develop more realistic RX models, the authors of \cite{akdeniz2018molecular} considered a partially absorbing RX and some recent studies, e.g., \cite{akkaya2014effect,ahmadzadeh2016comprehensive,sun2020expected,lotter2021saturating}, assumed a finite number of receptors uniformly distributed on the RX surface, where all receptors had the same size. Specifically, the authors of \cite{akkaya2014effect} assumed that molecules are absorbed once they hit a receptor, and the authors of \cite{ahmadzadeh2016comprehensive,sun2020expected} assumed that a reversible reaction occurs once a molecule hits a receptor. Very recently, the receptor occupancy induced by the competition of molecules when binding to the receptors of a postsynaptic neuron was investigated in \cite{lotter2021saturating}. While interesting, the previous studies may not be accurate when the receptors on the RX surface have different sizes and arbitrary locations, which may happen in practice. In fact, in nature, receptors tend to form clusters on the RX surface at specific locations \cite{duke2009equilibrium}. Since a given ligand can activate all receptors within a cluster, the cluster can be regarded as a single larger receptor. In \cite{lindsay2017first}, the authors investigated an RX covered by heterogeneous receptors with different sizes and arbitrary locations under steady-state conditions, but did not investigate the time-varying number of molecules absorbed by the RX, which is crucial in the context of MC.

As far as MC TX models are concerned, most previous studies assumed ideal point source TXs that release molecules instantaneously \cite{jamali2019channel}. This simplistic model ignores many important effects including the TX geometry, signaling pathways inside the TX, and chemical reactions during the release process, all of which affect practical transmission processes. Thus, the investigation of practical TX models can provide valuable insights into the design and implementation of end-to-end MC systems. Motivated by the exocytosis process of cells in nature \cite{jahn1999membrane}, we proposed a spherical TX model, namely a membrane fusion (MF)-based TX, as shown in Fig. \ref{MT}, in our recent work \cite{huang2021fusion}. This model assumes that molecules are stored within vesicles that are generated within the TX. Once the vesicles are instantaneously released from the center of the TX, they diffuse randomly inside the TX until they arrive at the TX membrane, where they can fuse with the membrane to release the encapsulated molecules. To the best knowledge of the authors, the signal transmission between MF-based TXs and RXs covered by heterogeneous receptors has not been studied, yet. 

In this paper, we consider an RX covered by heterogeneous receptors. Specifically, we model the receptors as absorbing patches (APs) and assume that there are multiple non-overlapping APs on the RX surface, where the APs can have different sizes and arbitrary but fixed locations. When molecules hit an AP, they are absorbed by the RX. Moreover, we consider two different TX models, namely, the point TX and the MF-based TX. We assume that the point TX and the center of the MF-based TX have a fixed distance to the center of the RX, as shown in Fig. \ref{sys}, and can be located at any point on the surface of the sphere defined by the TX-RX distance. Furthermore, we assume that the molecules may degrade while propagating in the channel. For the considered MC systems, we investigate the expected end-to-end channel impulse response (CIR) between TX and RX, averaged over all possible locations of the TX. Here, the expected CIR is defined as the expected molecule hitting rate at the RX \cite{jamali2019channel}. Additionally, we consider a sequence of bits transmitted from the TX to the RX and investigate the error probability of the end-to-end MC system. In summary, our major contributions are as follows:
\begin{itemize}
	\item[1)] We derive the expected molecule hitting rate, the expected fraction of absorbed molecules, and the expected asymptotic fraction of absorbed molecules as time approaches infinity at an RX with heterogeneous APs, which we refer to as AP-based RX. All expressions are functions of the sizes and locations of all APs. The derived expressions allow us to investigate the impact of different numbers, sizes, and locations of APs on molecular absorption.
	\item[2)] Aided by particle-based simulations (PBSs), we verify the accuracy of the derived analytical results. We also show that, for a given total area of the RX surface covered by APs, the expected number of absorbed molecules increases with the number of APs, but this increase becomes slower for larger numbers of APs.
	\item[3)] We compare three spatial distributions of the APs by examining the number of absorbed molecules at the RX. Specifically, we consider APs that are respectively evenly and randomly distributed across the entire RX surface, and APs that are evenly distributed within a region of the RX surface. Our results show that evenly distributed APs across the entire RX surface (i.e., the APs are equally spaced) result in a larger number of absorbed molecules than the other two distributions.
	\item[4)] We compare three different combinations of TX and RX models in terms of their CIRs and error probabilities. In particular, we consider the combination of i) a point TX and a fully absorbing RX (PTFR), ii) a point TX and an AP-based RX (PTAR), and iii) an MF-based TX and an AP-based RX (MTAR). Our results show that compared with the ideal PTFR model, AP-based RXs cause a lower peak and shorter tail of the CIR. Furthermore, both PTAR and MTAP result in a much higher error probability compared to PTFR. This indicates the importance of our analysis of realistic TX and RX models since the existing CIR and bit error rate (BER) analyses for ideal models are not accurate in practice.
\end{itemize}

This paper extends our preliminary work in \cite{huang2021analysis} as follows. First, we also consider MF-based TXs, while \cite{huang2021analysis} studied only point TXs. Second, we compare AP distributions for APs of identical and different sizes, while \cite{huang2021analysis} considered only APs of identical sizes. Third, we analyze the error probability of the end-to-end MC system. Fourth, we compare three different TX-RX models in terms of their CIRs and error probabilities.

The rest of this paper is organized as follows. In Section \ref{sm}, we describe the system model. In Section \ref{cirp}, we derive the CIR between a point TX and an AP-based RX. In Section \ref{ac}, we analyze the CIR between an MF-based TX and an AP-based RX. In Section \ref{oaacp}, we calculate the error probability of the MC system. In Section \ref{nr}, we present and discuss our numerical results. In Section \ref{con}, we draw some conclusions.

\section{System Model}\label{sm}
\begin{figure*}[!t]
	\centering
	
	\subfigure[Point TX]{
		\begin{minipage}[t]{0.5\linewidth}
			\centering
			\includegraphics[width=2.5in]{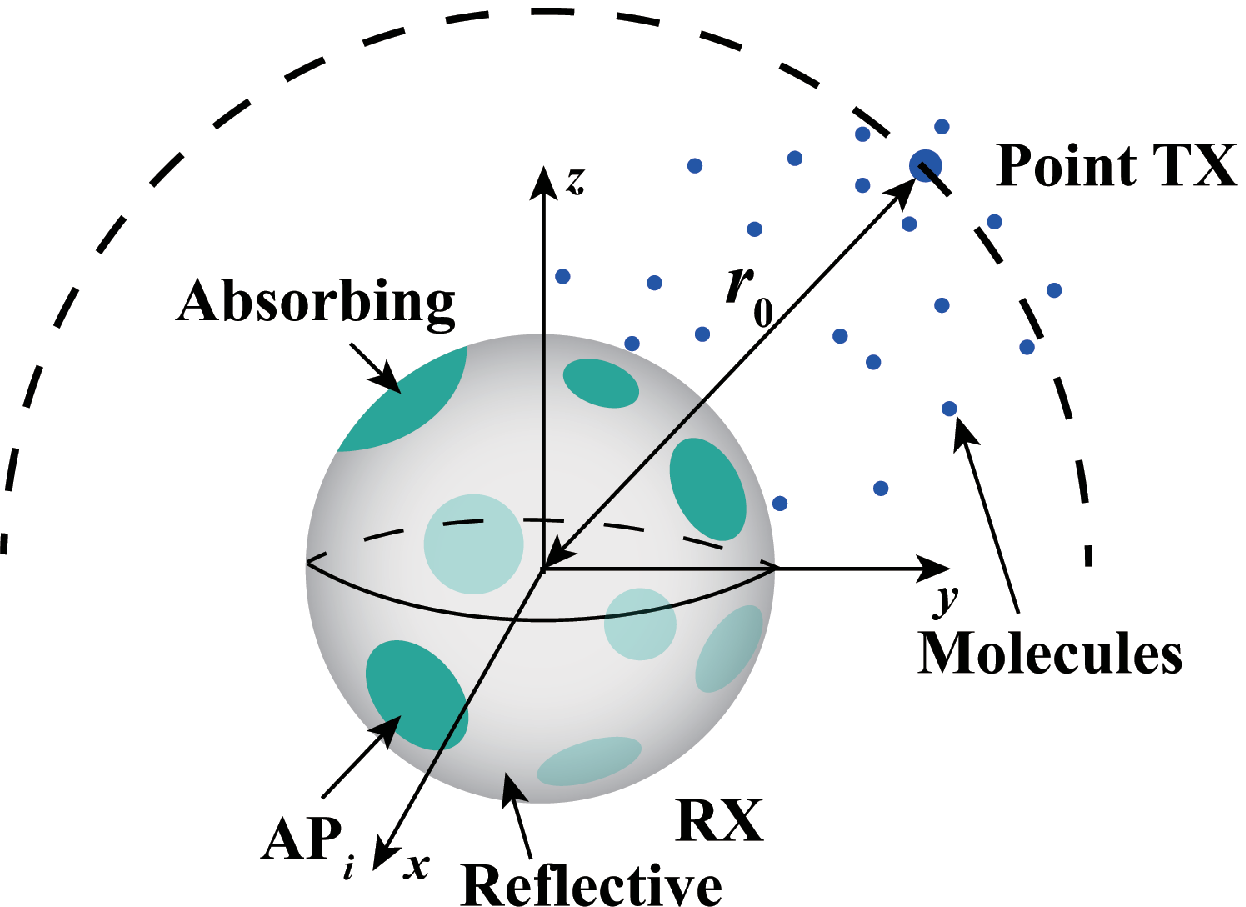}
			\label{PT}
		\end{minipage}%
	}%
	\subfigure[MF-based TX]{
		\begin{minipage}[t]{0.5\linewidth}
			\centering
			\includegraphics[width=3.4in]{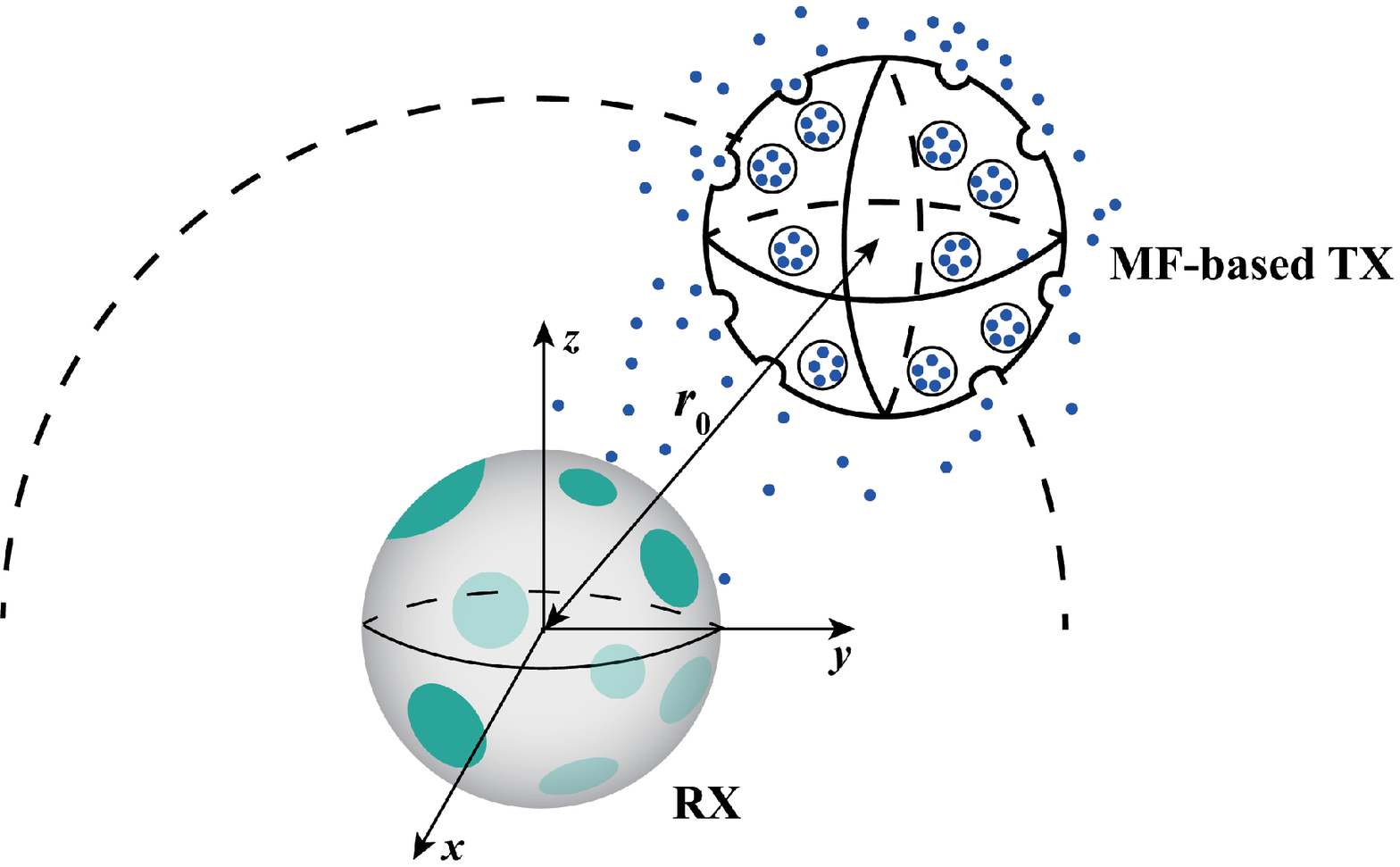}
			\label{MT}
		\end{minipage}%
	}
	\centering
	\caption{Illustration of the considered MC system model. (a): A point TX communicates with a spherical AP-based RX. (b): A spherical MF-based TX communicates with a spherical AP-based RX.}\label{sys}
\end{figure*}

In this paper, we consider an unbounded three-dimensional (3D) environment, where a TX communicates with a spherical RX, as depicted in Fig. \ref{sys}. We choose the center of the RX as the origin of the coordinate system and denote the radius of the RX by $r_{\ss\mathrm{R}}$. We assume that the propagation environment between TX and RX is a fluid medium with uniform temperature and viscosity. Once type $\sigma$ molecules are released from the TX, they diffuse randomly with a constant diffusion coefficient $D_\sigma$. We consider unimolecular degradation in the propagation environment, where the type $\sigma$ molecules can degrade to type $\hat{\sigma}$ molecules that cannot be recognized by the RX, i.e., $\sigma\stackrel{k_\mathrm{d}}{\longrightarrow}\hat{\sigma}$ \cite[Ch. 9]{chang2005physical}, where $k_\mathrm{d}\;[\mathrm{s}^{-1}]$ is the degradation rate. In the following, we first describe the novel AP-based RX model proposed in this paper and then present the considered TX models.

\subsection{RX Model}
We assume that there are $N_\mathrm{p}$ non-overlapping APs on the RX surface, and denote the $i$th AP by $\mathrm{AP}_i$. We approximate the shape of each AP as a circle, where the radius of $\mathrm{AP}_i$ is denoted by $a_i$. We further define $\mathcal{A}$ as the ratio of the total area of APs to the area of the RX surface, i.e., $\mathcal{A}=\sum_{i=1}^{N_\mathrm{p}}a_i^2/(4r_{\ss\R}^2)$. In spherical coordinates, we denote $\vec{l}_{i}=[r_{\ss\R}, \theta_i, \varphi_i]$ as the location of the center of $\mathrm{AP}_i$, where $\theta_i$ and $\varphi_i$ are the polar angle and azimuthal angle, respectively. As the APs are non-overlapping, the locations and radii of the APs satisfy $|\vec{l}_i-\vec{l}_j|\geq a_i+a_j$, $\forall i,j\in\{1,2,...,N_\mathrm{p}\}$, $i\neq j$. Once a molecule hits an AP, it is absorbed by the RX. As is customary in the literature \cite{akkaya2014effect}, we ignore the occupancy of the APs by molecules such that several molecules can be concurrently absorbed by the same AP. Furthermore, we model the area of the RX surface that is not covered by APs as perfectly reflective, which means that the molecules are reflected back once they hit this part of the RX surface. 
\subsection{TX Model}
In this paper, we consider the following two TX models:
\subsubsection{Point TX}\label{pt}
First, we consider a point TX communicating with an AP-based RX, cf. Fig. \ref{PT}. The point TX has distance $r_0$ to the center of the RX. For a given $r_0$, the TX can be located at any point on the surface of a sphere with radius $r_0$ around the RX and the CIR depends on the exact location of the TX. This is due to the fact that the considered RX has a heterogeneous boundary condition where the locations and sizes of the APs are determined by $\vec{l}_i$ and $a_i$, respectively. Thus, we focus on the expected CIR averaged over all possible locations of the TX with distance $r_0$ to the RX. We emphasize that the expected CIR for the given TX-RX distance is an important characteristic since, in practice, measuring the distance between two cells is much easier than determining their relative orientation. Specifically, cell $A$ can measure its distance to cell $B$ by detecting the nearby concentration of molecules released by cell $B$ \cite{lander2013cells}. 

\subsubsection{MF-based TX}
Second, we consider a spherical TX, namely the MF-based TX, communicating with an AP-based RX, cf. Fig. \ref{MT}. The MF-based TX model was proposed in our recent work \cite{huang2021fusion}, and employs MF between vesicles generated within the TX and the TX membrane to release the molecules encapsulated in the vesicles. We denote $r_{\ss\T}$ as the radius of the MF-based TX. Each vesicle contains and transports $\eta$ molecules of type $\sigma$. Once vesicles are instantaneously released from the center of the MF-based TX, they diffuse randomly with a constant diffusion coefficient $D_\mathrm{v}$ within the TX. We model the MF process between the vesicle and the TX membrane as an irreversible reaction with forward reaction rate $k_\mathrm{f}\;[\mu\mathrm{m}/\mathrm{s}]$. After MF, the molecules stored by the vesicle are instantaneously released into the propagation environment. We assume that the spherical TX does not impede the random diffusion of the molecules in the propagation environment. As shown in \cite{huang2021fusion}, this assumption is valid as long as $r_{\ss\T}+r_{\ss\R}<0.9r_0$. Furthermore, similar to Section \ref{pt}, we assume that the center of the MF-based TX is located at a distance $r_0$ from the center of the RX. Thus, the center of the MF-based TX can be located on any point of the surface of a sphere with radius $r_0$ around the RX, and the expected CIR is averaged over all possible locations of the MF-based TX.

\subsection{Bit Transmission}\label{bt}
Information sent from the TX to the RX is encoded into a sequence of $Q$ binary bits, denoted by $\mathbf{b}_{1:Q}=[b_1, b_2, ..., b_Q]$, where $b_q$ is the $q$th bit, $q\in\{1,\ldots,Q\}$. We denote $T_\mathrm{b}$ as the bit interval length and assume that $b_1$ is transmitted at time $t=0$. We also assume that each bit is transmitted with probabilities $\mathrm{Pr}(b_q=0)=P_0$ and $\mathrm{Pr}(b_q=1)=P_1=1-P_0$, where $\mathrm{Pr}(\cdot)$ denotes probability. We adopt ON/OFF keying for modulation. Thus, $N_\sigma$ molecules or $N_\mathrm{v}$ vesicles are instantaneously released at the beginning of the bit interval to transmit bit 1 for a point TX and an MF-based TX, respectively, while no molecules or vesicles are released to transmit bit 0. We further assume that the TX and RX are perfectly synchronized by reliable MC synchronization schemes, e.g., those proposed in \cite{jamali2017symbol}. To demodulate $b_q$ at the RX, we adopt the widely-used threshold detector that compares the number of absorbed molecules during the $q$th bit interval with a decision threshold \cite{kuran2020survey}.

\section{Analysis of the Channel Impulse Response for a Point TX}\label{cirp}
In this section, we assume a point TX and derive i) the expected molecule hitting rate, ii) the expected fraction of absorbed molecules, and iii) the expected asymptotic fraction of absorbed molecules as $t\rightarrow\infty$ at the AP-based RX by applying boundary homogenization \cite{dagdug2016boundary}. To this end, we first derive the expected CIR of an RX with uniform surface reaction rate. Here, a uniform surface reaction rate implies that the reactivity of the molecules is identical for all points on the RX surface. We then consider the system in steady state and derive the diffusion current of molecules across the RX surface. We note that the diffusion current $[\textrm{molecule}\cdot\textrm{s}^{-1}]$ is the rate of the molecule movement per unit time. Based on the diffusion current in steady state, we finally determine the effective reaction rate and apply it to derive the expected CIR of an AP-based RX.
\subsection{Problem Formulation}
In spherical coordinates, we denote the molecule concentration at time $t$ at location $\vec{r}$ by $C(\vec{r},t)$, $|\vec{r}|\geq r_{\ss\R}$. For impulsive release of molecules from the TX at time $t=0$, the initial condition can be expressed as follows\cite[Eq. (3.61)]{schulten2000lectures}
\begin{align}\label{ic}
C(\vec{r},t\rightarrow 0)=\frac{1}{4\pi r_0^2}\delta(|\vec{r}|-r_0),
\end{align}
where $\delta(\cdot)$ is the Dirac delta function. After their release, the movement of the molecules by diffusion in the propagation environment is described by Fick's second law as follows \cite{berg1993random}
\begin{align}\label{cr}
\frac{\partial C(\vec{r}, t)}{\partial t}=D_\sigma\nabla^2C(\vec{r}, t)-k_\mathrm{d}C(\vec{r}, t),
\end{align}
where $\nabla^2$ is the 3D spherical Laplacian. For molecules hitting the RX surface, the reaction between the molecules and the RX surface is described by the radiation boundary condition \cite[Eq. (3.64)]{schulten2000lectures}, given by
\begin{align}\label{bc}
D_\sigma\frac{\partial C(|\vec{r}|, t)}{\partial |\vec{r}|}\bigg|_{|\vec{r}|=r_{\ss\R}}=wC(r_{\ss\R}, t),
\end{align}
where $w\;[\mu\mathrm{m}/\mathrm{s}]$ denotes the reaction rate. We note that $w\rightarrow\infty$ when $\vec{r}\in\Omega_\mathrm{AP}$ while $w=0$ when $\vec{r}\in\Omega_\mathrm{R}$, where $\Omega_\mathrm{AP}$ and $\Omega_\mathrm{R}$ denote the areas of the RX surface that are covered and not covered by APs, respectively. Due to the heterogeneous boundary condition in \eqref{bc}, it is difficult to directly solve \eqref{cr} to obtain $C(\vec{r},t)$. Hence, in this paper, we apply boundary homogenization to derive the expected CIR. The main idea of boundary homogenization is to replace the heterogeneous boundary condition in \eqref{bc} by a uniform boundary condition with an appropriately chosen effective surface reaction rate, denoted by $w_\mathrm{e}$. This means that we replace the RX surface with heterogeneous boundary conditions shown in Fig. \ref{sys} with an equivalent uniform surface with reaction rate $w_\mathrm{e}$. Hence, \eqref{bc} can be reexpressed by replacing $w$ with $w_\mathrm{e}$. We derive the expected CIR of the RX and $w_\mathrm{e}$ in the following subsections.

\subsection{Expected CIR of RX with Uniform Surface Reaction Rate}
In this subsection, we analyze the expected CIR of an RX with uniform surface reaction rate $w$. Based on the initial condition in \eqref{ic} and the boundary condition in \eqref{bc}, the authors of \cite{schulten2000lectures}\footnote{We note that the RX in \cite{schulten2000lectures} is fully covered by receptors, where the forward reaction rate between receptors and molecules is given by $w$.} derived $C(|\vec{r}|, t)$ by solving \eqref{cr} when $k_\mathrm{d}=0$. We denote $h_\mathrm{u}(t, w)$ as the expected molecule hitting rate at an RX with uniform surface reaction rate. We derive $h_\mathrm{u}(t, w)$ including the effect of molecule degradation in the following lemma.
\begin{lemma}
	The expected molecule hitting rate at an RX with uniform surface reaction rate $w$ at time $t$ when molecules are released from a point TX at time $t=0$ is given by
	\begin{align}\label{hu}
	&h_\mathrm{u}(t,w)=\frac{r_{\ss\mathrm{R}}w}{r_0}\left[\frac{1}{\sqrt{\pi D_\sigma t}}\exp\left(-\frac{\varepsilon^2}{t}-k_\mathrm{d}t\right)-\gamma(w)\right.\notag\\&\times\left.\!\!\exp\!\left[\gamma(w)(r_0-r_{\ss\mathrm{R}})\!+\!\zeta(w)t\right]
	\mathrm{erfc}\!\left(\!\!\frac{\varepsilon}{\sqrt{t}}\!+\!\gamma(w)\sqrt{D_\sigma t}\right)\!\!\right],
	\end{align}
	where $\varepsilon\!=\!\frac{r_0-r_{\ss\mathrm{R}}}{\sqrt{4D_\sigma}}$, $\gamma(w)\!=\!\frac{wr_{\ss\mathrm{R}}+D_\sigma}{D_\sigma r_{\ss\mathrm{R}}}$,  $\zeta(w)\!=\!\gamma(w)^2D_\sigma-k_\mathrm{d}$, and $\mathrm{erfc}(\cdot)$ is the complementary error function \cite{andrews1998special}.
	\begin{IEEEproof}
		According to \cite{heren2015effect}, $h_\mathrm{u}(t,w)$ can be obtained as $h_\mathrm{u}(t,w)=h_\mathrm{u}(t,w)\big|_{k_\mathrm{d}=0}\times\exp\left(-k_\mathrm{d}t\right)$. We also find $h_\mathrm{u}(t,w)\big|_{k_\mathrm{d}=0}=4\pi r_{\ss\R}^2wC(r_{\ss\R}, t)\big|_{k_\mathrm{d}=0}$ based on \cite[Eq. (3.107)]{schulten2000lectures}. Combining these two results, we obtain \eqref{hu}.
	\end{IEEEproof}
\end{lemma}

We now denote $H_\mathrm{u}(t,w)$ as the fraction of molecules captured by the RX by time $t$ and present it in the following corollary.
\begin{corollary}
	The fraction of molecules captured by an RX with uniform surface reaction rate $w$ by time $t$, when molecules are released from a point TX, is given by
	\begin{align}\label{Hu}
	H_\mathrm{u}(t,w)=\frac{r_{\ss\mathrm{R}}w}{r_0}\left[\alpha_1(t)-\alpha_2(t,w)\right],
	\end{align}
	where
	\begin{align}\label{d1}
	\alpha_1(t)=&\frac{1}{2\sqrt{k_\mathrm{d}D_\sigma}}\left[\exp(-\beta)\mathrm{erfc}\left(\frac{\varepsilon}{\sqrt{t}}-\sqrt{k_\mathrm{d}t}\right)\right.\notag\\&\left.-\exp\left(\beta\right)\mathrm{erfc}\left(\frac{\varepsilon}{\sqrt{t}}+\sqrt{k_\mathrm{d}t}\right)\right]
	\end{align}
	and
	\begin{align}\label{d2}		\alpha_2(t,w)\!=\!\frac{1}{2\zeta(w)}\!\left[\psi_1(t,w)\!-\!\psi_2(t,w)\right]-\frac{\gamma(w)\exp(-\beta)}{\zeta(w)}
	\end{align}
	with
	\begin{align}\label{e1}		\psi_1(t,w)=&2\gamma(w)\exp\left(\gamma(w)(r_0-r_{\ss\mathrm{R}})+\zeta(w)t\right)\notag\\&\times\mathrm{erfc}\left(\frac{\varepsilon}{\sqrt{t}}+\gamma(w)\sqrt{D_\sigma t}\right),
	\end{align}
	\begin{align}\label{e2}		\psi_2(t,w)\!=&\!\left(\!\gamma(w)^2\sqrt{\!\frac{D_\sigma}{k_\mathrm{d}}}\!-\!\gamma(w)\!\right)\exp(-\beta)\mathrm{erf}\left(\frac{\varepsilon}{\sqrt{t}}-\sqrt{k_\mathrm{d}t}\right)\notag\\
	&-\left(\gamma(w)^2\sqrt{\frac{D_\sigma}{k_\mathrm{d}}}
	+\gamma(w)\right)\bigg[\exp(-\beta)-\exp(\beta)\notag\\&\times\mathrm{erfc}\left(\frac{\varepsilon}{\sqrt{t}}
	+\sqrt{k_\mathrm{d}t}\right)\bigg],
	\end{align}
	$\mathrm{erf}(\cdot)=1-\mathrm{erfc}(\cdot)$ is the error function \cite{andrews1998special}, and $\beta=(r_0-r_{\ss\mathrm{R}})\sqrt{k_\mathrm{d}/D_\sigma}$.
\end{corollary}
\begin{IEEEproof}
	$H_\mathrm{u}(t,w)$ can be obtained from $H_\mathrm{u}(t,w)=\int_{0}^{t}h_\mathrm{u}(u,w)\mathrm{d}u$. By substituting \eqref{hu} into this equation, we obtain \eqref{Hu}.
\end{IEEEproof}

We next denote $H_{\mathrm{u}, \infty}(w)$ as the asymptotic fraction of molecules captured by the RX as $t\rightarrow\infty$. We present $H_{\mathrm{u}, \infty}(w)$ in the following corollary.
\begin{corollary}\label{asy}
	When molecules are released from a point TX, as $t\rightarrow\infty$, the asymptotic fraction of molecules captured by an RX with uniform surface reaction rate $w$ is given by
	\begin{align}\label{Ha}
	H_{\mathrm{u},\infty}(w)=\frac{r_{\ss\mathrm{R}}
		w\left(\gamma(w)-\sqrt{\frac{k_\mathrm{d}}{D_\sigma}}\right)}{r_0\zeta(w)}\exp(-\beta).
	\end{align}
	\begin{IEEEproof}
		Please see Appendix \ref{AA}.
	\end{IEEEproof}
\end{corollary}

\subsection{Effective Reaction Rate}
In this subsection, we determine the effective reaction rate $w_\mathrm{e}$. First, we investigate the diffusion flux and diffusion current of the molecules, which lays the foundation for deriving $w_\mathrm{e}$. We denote $J$ as the diffusion flux $[\textrm{molecule}\cdot\textrm{m}^{-2}\cdot\textrm{s}^{-1}]$ of the molecules, which is the rate at which the molecules move across a unit area in a unit time \cite{berg1993random}. We note that $J$ is given by \cite[Eq. (2.6)]{berg1993random}
\begin{align}
J=-D_\sigma\frac{\partial C(|\vec{r}|, t)}{\partial |\vec{r}|}\bigg|_{|\vec{r}|=r_{\ss\R}}.
\end{align}
We next denote $I$ as the diffusion current of the molecules, which represents the rate of molecule movement across the RX surface in a unit time. We note that $I$ is given by $I=-4\pi r_{\ss\mathrm{R}}^2J$. The negative sign is due to the inward diffusion current. In the steady state, the partial derivative of the molecule concentration with respect to time is zero. Thus, we have $\partial C(\vec{r},t)/\partial t=0$. If we set $k_\mathrm{d}=0$ in \eqref{cr}, we obtain
\begin{align}\label{nc}
\nabla^2C(\vec{r}, t)=0.
\end{align}

As explained in \cite{berg1977physics}, since \eqref{nc} is analogous to the Laplace's equation for the electrostatic potential in charge-free space, the diffusion current to an isolated absorbing RX of any size and shape can be expressed as \cite[Eq. (2.24)]{berg1993random}
\begin{align}\label{I}
I=4\pi D_\sigma GC_0,
\end{align}
where we define $G$ as the ``capacitance" of the RX and $C_0$ is the molecule concentration at $|\vec{r}|\rightarrow\infty$ in steady state. We note that $C_0=1$ was adopted in some previous studies, e.g., \cite{lindsay2017first,ahmadzadeh2016comprehensive}. We further note that $G$ measures the ability of the MC RX to absorb molecules, which is different from the conventional electrical capacitance of a conductor, denoted by $\hat{G}$. However, according to \cite{berg1993random}, if the MC RX and the conductor have the same size and shape, $G$ can be obtained as $G=\hat{G}/(4\pi\epsilon_0)$, where $\epsilon_0$ is the vacuum permittivity \cite{chung1982electrical}. Since expressions for $\hat{G}$ have been derived for a variety of conductors, we can obtain the corresponding $G$ directly. For example, we have $\hat{G}=4\pi\epsilon_0r_{\ss\R}$ for a spherical conductor with radius $r_{\ss\R}$ \cite[Eq. (1)]{crowley2008simple}. According to the relationship between $G$ and $\hat{G}$, we obtain the ``capacitance" of a fully absorbing RX with the same radius, denoted by $G_\mathrm{a}$, as $G_\mathrm{a}=r_{\ss\R}$. Therefore, the diffusion current of a fully absorbing RX, denoted by $I_\mathrm{a}$, is given by $I_\mathrm{a}=4\pi D_\sigma r_{\ss\mathrm{R}}C_0$, which aligns with the derivation in \cite[Eq. (1)]{berg1977physics} based on solving \eqref{nc}.
For the AP-based RX, the ``capacitance" of the RX, denoted by $G_\mathrm{p}$, was derived in \cite{lindsay2017first} by using the method of matched asymptotic expansions. Specifically, $G_\mathrm{p}$ is a function of the size and location of each AP, and given by \cite[Eq. (3.37a)]{lindsay2017first}
\begin{align}\label{Gc}
&\frac{1}{G_\mathrm{p}}=\frac{2}{N_\mathrm{p}\overline{m}\kappa r_{\ss\mathrm{R}}}\Bigg[1+\frac{\kappa}{2N_\mathrm{p}\overline{m}}
\ln\left(\frac{\kappa}{2}\right)\sum_{i=1}^{N_\mathrm{p}}m_i^2
+\frac{\kappa}{N_\mathrm{p}\overline{m}}\notag\\
&\times\bigg(\sum_{i=1}^{N_\mathrm{p}}m_is_i+2\sum_{i=1}^{N_\mathrm{p}}
\sum_{j=i+1}^{N_\mathrm{p}}m_im_j\mathcal{F}(\vec{l}_i',\vec{l}_j')\bigg)\!
+\!\left(\kappa\ln\left(\frac{\kappa}{2}\right)\!\!\right)^2\notag\\
&\times\frac{\vartheta}{4N_\mathrm{p}\overline{m}}
+\mathcal{O}\left(\kappa^2\ln\left(\frac{\kappa}{2}\right)\right)\Bigg],
\end{align}
where $\kappa=\frac{a_1}{r_{\ss\mathrm{R}}}$, $m_i=\frac{2a_i}{r_{\ss\mathrm{R}}\kappa\pi}$, $\overline{m}=\frac{1}{N_\mathrm{p}}\sum_{i=1}^{N_\mathrm{p}}m_i$, $s_i=\frac{m_i}{2}\left(\ln\left(\frac{4a_i}{r_{\ss\mathrm{R}}\kappa}\right)-\frac{3}{2}\right)$, $\vartheta=\frac{\left(\sum_{i=1}^{N_\mathrm{p}}m_i^2\right)^2}{N_\mathrm{p}\overline{m}}-\sum_{i=1}^{N_\mathrm{p}}m_i^3$, and
\begin{align}\label{F}
\mathcal{F}(\vec{l}_i', \vec{l}_j')=\frac{1}{|\vec{l}_i'-\vec{l}_j'|}+\frac{1}{2}\ln|\vec{l}_i'\!-\!\vec{l}_j'|\!-\!\frac{1}{2}\!\ln\!\left(2+|\vec{l}_i'-\vec{l}_j'|\right)
\end{align}
with $\vec{l}_i'=\vec{l}_i/r_{\ss\mathrm{R}}$. In \eqref{Gc}, $\mathcal{O}(\cdot)$ represents the infinitesimal of higher order, which is omitted during calculation. The accuracy of \eqref{Gc} when the $\mathcal{O}(\cdot)$ term is ignored will be discussed in Fig. \ref{ga} in Section \ref{nrpt}.

When all APs have the same size, \eqref{Gc} can be further simplied as shown in the following corollary.
\begin{corollary}
	When all APs have identical sizes, i.e., $a_1=a_2=\cdots=a_{\ss{N_\mathrm{p}}}$, $G_\mathrm{p}$ can be simplified as follows
\begin{align}\label{G1}
\frac{1}{G_\mathrm{p}}=&\frac{\pi}{N_\mathrm{p}\kappa r_{\ss\mathrm{R}}}\Bigg[1+\frac{\kappa}{\pi}\bigg(\!\ln(2\kappa)
-\frac{3}{2}+\frac{4}{N_\mathrm{p}}\sum_{i=1}^{N_\mathrm{p}}\sum_{j=i+1}^{N_\mathrm{p}}
\mathcal{F}(\vec{l}_i',\vec{l}_j')\bigg)\notag\\
&+\mathcal{O}\left(\kappa^2\ln\left(\frac{\kappa}{2}\right)\right)\Bigg].
\end{align}
\end{corollary}
\begin{IEEEproof}
	When $a_1=a_2=\cdots=a_{\ss{N_\mathrm{p}}}$, we have $m_i=\frac{2}{\pi}$ and $\vartheta=0$. By substituting $m_i$ and $\vartheta$ into \eqref{Gc}, we obtain \eqref{G1}.
\end{IEEEproof}

When the APs have the same size and are evenly distributed, i.e., equally spaced, on the RX surface, \eqref{G1} can be further simplified as $N_\mathrm{p}\rightarrow\infty$ by using the mean-field approximation method \cite{cheviakov2010asymptotic}, which results in \cite[Eq. (4.7)]{lindsay2017first}
\begin{align}\label{Gnl}
\frac{1}{G_\mathrm{p}}\approx&\frac{1}{r_{\ss\R}}\left(1+\frac{\pi}{N_\mathrm{p}\kappa}+\frac{\frac{1}{2}\ln(\kappa\sqrt{N_\mathrm{p}})+\ln 2-\frac{3}{2}}{N_\mathrm{p}}-2N_\mathrm{p}^{-\frac{1}{2}}\right.\notag\\&\left.+N_\mathrm{p}^{-\frac{3}{2}}\right).
\end{align}

We note that \eqref{Gnl} only depends on $N_\mathrm{p}$ and not on the exact locations of the APs. We will confirm the accuracy of \eqref{Gnl}  in Fig. \ref{df} in Section \ref{nrpt}. Our results in Fig. \ref{df} show that even though \eqref{Gnl} is derived as $N_\mathrm{p}\rightarrow\infty$, \eqref{Gnl} is still accurate when $N_\mathrm{p}$ is small.

The capacitance of an RX with a single AP can be obtained by setting $N_\mathrm{p}=1$ in \eqref{G1}. In \cite{lindsay2017first}, the authors applied a higher order asymptotic expansion to derive a more accurate expression, which is given by \cite[Eq. (6.31)]{lindsay2017first}
\begin{align}\label{g1a}
\frac{1}{G_\mathrm{p}}\bigg|_{N_\mathrm{p}=1}=&\frac{\pi}{\kappa r_\mathrm{R}}\bigg[1+\frac{\kappa}{\pi}\left(\ln(2\kappa)-\frac{3}{2}\right)-\frac{\kappa^2}{\pi^2}\left(\frac{\pi^2+21}{36}\right)\notag\\&+\mathcal{O}(\kappa^3\ln\kappa)\bigg].
\end{align}

\begin{table}[!t]
	\newcommand{\tabincell}[2]{\begin{tabular}{@{}#1@{}}#2\end{tabular}}
	\centering
	\caption{Summary of expressions for Calculating $\frac{1}{G_\mathrm{p}}$:}\label{tab2}
    \begin{tabular}{|c|c|}
			\hline
			\textbf{Expression}&\textbf{Size and Distribution of APs}\\
			\hline
			\eqref{Gc}& Any size, any distribution\\
			\hline
			\eqref{G1}& Identical sizes, any distribution\\
			\hline
			\eqref{Gnl}& Identical sizes, evenly distributed\\
			\hline
			\eqref{g1a}& Single AP\\
			\hline
			
		\end{tabular}
\end{table} 

In Table \ref{tab2}, we summarize the expressions for calculating $1/G_{\mathrm{p}}$ for different AP sizes and distributions. We denote $h_\mathrm{p}(t)$, $H_\mathrm{p}(t)$, and $H_{\mathrm{p},\infty}$ as the expected molecule hitting rate, the expected fraction of absorbed molecules, and the expected asymptotic fraction of absorbed molecules at an AP-based RX when the molecules are released from a point TX, respectively. We present $h_\mathrm{p}(t)$, $H_\mathrm{p}(t)$, and $H_{\mathrm{p},\infty}$ along with $w_\mathrm{e}$ in the following theorem.
\begin{theorem}
	The expected molecule hitting rate, the expected fraction of absorbed molecules, and the expected asymptotic fraction of absorbed molecules as $t\rightarrow\infty$ at an AP-based RX when the molecules are released at the TX at time $t=0$ are given by
	\begin{align}\label{com}
	\!\!\!h_\mathrm{p}\!(t)\!=\!h_\mathrm{u}\!(t,w_\mathrm{e}), H_\mathrm{p}(t)\!=\!H_\mathrm{u}(t,w_\mathrm{e}), H_{\mathrm{p},\infty}\!=\!H_{\mathrm{u},\infty}(w_\mathrm{e}),
	\end{align}
	respectively, where $h_\mathrm{u}(t,w)$, $H_\mathrm{u}(t,w)$, and $H_{\mathrm{u}, \infty}(w)$ are given by \eqref{hu}, \eqref{Hu}, and \eqref{Ha}, respectively, and
	\begin{align}\label{wee}
	w_\mathrm{e}=\frac{D_\sigma G_\mathrm{p}}{r_{\ss\mathrm{R}}(r_{\ss\mathrm{R}}-G_\mathrm{p})}.
	\end{align}
	The expressions for $G_\mathrm{p}$ are summarized in Table \ref{tab2}.
\end{theorem}
\begin{IEEEproof}
	We note that the diffusion current of molecules across an AP-based RX, denoted by $I_\mathrm{p}$, can be obtained by replacing $G$ with $G_\mathrm{p}$ in \eqref{I}. The ratio between the expected asymptotic fraction of absorbed molecules at an AP-based RX and at a fully absorbing RX is equal to the ratio between the corresponding diffusion currents \cite{ahmadzadeh2016comprehensive}. Thus, we have \begin{align}\label{ra}
	\frac{H_{\mathrm{u},\infty}(w_\mathrm{e})|_{k_\mathrm{d}=0}}
	{H_{\mathrm{a},\infty}}=\frac{I_\mathrm{p}}{I_\mathrm{a}},
	\end{align}
	where $H_{\mathrm{u},\infty}(w_\mathrm{e})\big|_{k_\mathrm{d}=0}
	=r_{\ss\mathrm{R}}^2w_\mathrm{e}/(r_0(w_\mathrm{e}r_{\ss\mathrm{R}}+D_\sigma))$ based on \eqref{Ha}, and $H_{\mathrm{a},\infty}=r_{\ss\mathrm{R}}/r_0$ is the asymptotic fraction of absorbed molecules as $t\rightarrow\infty$ at a fully absorbing RX \cite[Eq. (3.116)]{schulten2000lectures}. By solving \eqref{ra}, we obtain \eqref{wee}. By substituting $w_\mathrm{e}$ into \eqref{hu}, \eqref{Hu}, and \eqref{Ha}, we obtain \eqref{com}.
\end{IEEEproof}

\section{Analysis of the Channel Impulse Response for an MF-Based TX}\label{ac}
In this section, we consider an MF-based TX and derive the i) expected molecule hitting rate, ii) expected fraction of molecules absorbed, and iii) expected asymptotic fraction of molecules absorbed as time $t\rightarrow\infty$ at an AP-based RX. To this end, we first derive the expected molecule hitting rate at the RX when molecules are uniformly released from a spherical TX membrane with radius $r_{\ss\T}$ at time $t=0$. A uniform release of molecules implies that the likelihood that a molecule is released is identical for any point on the spherical TX surface. We define the molecule release rate of an MF-based TX, denoted by $f_\mathrm{r}(t)$, as the probability that one molecule is released during the time interval $[t, t+\delta t]$ from the TX membrane, when the vesicle storing this molecule is released from the center of the TX at time $t=0$. Here, $\delta t$ represents an infinitesimally small time interval. Then,  using $f_\mathrm{r}(t)$ given in \cite[Eq. (5)]{huang2021fusion} and the derived expected hitting rate, we derive the expected CIR between the MF-based TX and the AP-based RX.

\subsection{Expected Hitting Rate at AP-based RX for Uniform Release of Molecules}
The expected CIR analysis of the point TX in Section \ref{cirp} lays the foundation for deriving the expected CIR of the MF-based TX. As vesicles are released from the center of the MF-based TX, molecules are uniformly released from the TX membrane. Therefore, to derive the expected end-to-end CIR between the MF-based TX and the RX, we first calculate the expected CIR when molecules are uniformly released from the TX membrane at time $t=0$. We denote the expected hitting rate at the RX due to the uniform release of molecules as $h_{\mathrm{s}}(t)$ and present $h_\mathrm{s}(t)$ in the following lemma.

\begin{lemma}
	The expected molecule hitting rate at an AP-based RX at time $t$ when a spherical TX uniformly releases molecules from its surface at time $t=0$ is given by
	\begin{align}\label{hs}
	h_\mathrm{s}(t)\!&=\!\frac{r_{\ss\R}w_\mathrm{e}}{2r_{\ss\T}r_0}\!\left[\xi_1(t, r_0\!-\!r_{\ss\T}-r_{\ss\R})-\xi_1(t, r_0+r_{\ss\T}-r_{\ss\R})\right],
	\end{align}
	where
	\begin{align}
		\xi_1(t,z)\!=\!\exp\left(\gamma(w_\mathrm{e})z+\zeta(w_\mathrm{e})t\right)\mathrm{erfc}\left(\varpi(w_\mathrm{e})\sqrt{t}+\frac{z}{\sqrt{4D_\sigma t}}\right)
	\end{align} 
	with $\varpi(w_\mathrm{e})=\gamma(w_\mathrm{e})\sqrt{D_\sigma}$.
\end{lemma}

\begin{IEEEproof}
	When molecules are released from an arbitrary point $\nu$ on the membrane of a spherical TX, the expected molecule hitting rate at the RX is obtained by replacing $r_0$ with $r_\nu$ in $h_\mathrm{p}(t)$ in \eqref{com}, where $r_\nu$ is the distance between point $\nu$ and the center of the RX. Following \cite[Appendix B]{huang2021fusion}, we derive $h_\mathrm{s}(t)$ by computing the surface integral of $h_\mathrm{p}(t, r_\nu)$ over the TX membrane, which is given by
	\begin{align}\label{hst}
	h_\mathrm{s}(t)=\frac{1}{2r_{\ss\T}}\int_{-r_{\ss\T}}^{r_{\ss\T}}h_\mathrm{p}(t,r_\nu)\big|_{r_\nu=\sqrt{r_{\ss\T}^2+r_0^2-2r_0x}}\mathrm{d}x.
	\end{align}
	
	By substituting $h_\mathrm{p}(t,r_\nu)$ given in \eqref{com} into \eqref{hst}, we obtain \eqref{hs}.
\end{IEEEproof}

We denote $H_\mathrm{s}(t)$ as the expected fraction of absorbed molecules at the RX and present $H_\mathrm{s}(t)$ in the following corollary.
\begin{corollary}
	The expected fraction of molecules absorbed at an AP-based RX by time $t$ when a spherical TX uniformly releases molecules from its surface at time $t=0$ is given by
	\begin{align}\label{Hst}
	H_\mathrm{s}(t)\!=\!\frac{r_{\ss\R}w_\mathrm{e}}{2r_{\ss\T}r_0\zeta(w_\mathrm{e})}\!\left[\xi_2(t,r_0\!-\!r_{\ss\T}-\!r_{\ss\R})\!-\xi_2(t, r_0+r_{\ss\T}-r_{\ss\R})\right],
	\end{align}
	where $\xi_2(t,z)$ is given in \eqref{xi} on the top of the next page.
	\begin{figure*}[!t]
		\begin{align}\label{xi}
		&\xi_2(t,z)=\exp\left(\gamma(w_\mathrm{e})z+\zeta(w_\mathrm{e})t\right)\mathrm{erfc}\left(\frac{z}{\sqrt{4D_\sigma t}}+\varpi(w_\mathrm{e})\sqrt{t}\right)-\frac{1}{2\sqrt{k_\mathrm{d}}}\exp\left(-z\sqrt{\frac{k_\mathrm{d}}{D_\sigma}}\right)\left[\left(\varpi(w_\mathrm{e})-\sqrt{k_\mathrm{d}}\right)\right.\notag\\&\left.\times\mathrm{erf}\!\left(\!\frac{z}{\sqrt{4D_\sigma t}}-\sqrt{k_\mathrm{d}t}\right)\!-\left(\varpi(w_\mathrm{e})+\sqrt{k_\mathrm{d}}\right)\left(1-\exp\left(2z\sqrt{\frac{k_\mathrm{d}}{D_\sigma}}\right)\mathrm{erfc}\left(\frac{z}{\sqrt{4D_\sigma t}}+\sqrt{k_\mathrm{d}t}\right)\right)\right]-\exp\left(-z\sqrt{\frac{k_\mathrm{d}}{D_\sigma}}\right).
		\end{align}
		\hrulefill \vspace*{-1pt}
	\end{figure*}
\end{corollary}

\begin{IEEEproof}
	$H_\mathrm{s}(t)$ can be obtained as $H_\mathrm{s}(t)=\int_{0}^{t}h_\mathrm{s}(u)\mathrm{d}u$. By substituting \eqref{hs} into this equation, we obtain \eqref{Hst}.
\end{IEEEproof}

\subsection{Expected CIR of AP-based RX for MF-Based TX}
In this subsection, we derive the expected CIR for an AP-based RX when the molecules are released from an MF-based TX. We denote $h_{\ss\mathrm{MF}}(t)$ as the expected molecule hitting rate at the RX at time $t$ when vesicles are released from the center of an MF-based TX at time $t=0$ and present $h_{\ss\mathrm{MF}}(t)$ in the following theorem.
\begin{theorem}\label{t2}
	The expected molecule hitting rate at an AP-based RX at time $t$ when vesicles are released from the center of an MF-based TX at time $t=0$ is given by
	\begin{align}\label{hmm}
	h_{\ss\mathrm{MF}}(t)&=\frac{2r_{\ss\T}r_{\ss\R}k_\mathrm{f}w_\mathrm{e}\exp(\zeta(w_\mathrm{e})t)}{r_0}\sum_{n=1}^{\infty}\frac{\lambda_n^3j_0(\lambda_nr_{\ss\T})}{2\lambda_nr_{\ss\T}-\sin(2\lambda_nr_{\ss\T})}\notag\\&\times\left[\xi_3(t,r_0-r_{\ss\T}-r_{\ss\R})-\xi_3(t, r_0+r_{\ss\T}-r_{\ss\R})\right],
	\end{align}
where $\xi_3(t,z)=\exp\left(\gamma(w_\mathrm{e})z\right)\varsigma_1(t,z)$ with
\begin{align}\label{var}
&\varsigma_1(t,z)=\int_{0}^{t}\exp\left(-(\zeta(w_\mathrm{e})+D_\mathrm{v}\lambda_n^2)u\right)\notag\\&\times\mathrm{erfc}\left(\varpi(w_\mathrm{e})\sqrt{t-u}+\frac{z}{\sqrt{4D_\sigma(t-u)}}\right)\mathrm{d}u, 
\end{align} 
$j_0(\cdot)$ is the zeroth order spherical Bessel function of the first kind \cite{olver1960bessel}, and $\lambda_n$ is obtained by solving 
\begin{align}\label{Dvl}
-D_\mathrm{v}\lambda_nj_0'\left(\lambda_nr_{\ss\T}\right)=k_\mathrm{f}j_0\left(\lambda_nr_{\ss\T}\right),
\end{align}
with $j_0'(z)=\frac{\mathrm{d} j_0(z)}{\mathrm{d} z}$ and $n=1,2,...$. We note that \eqref{var} can be numerically calculated by the built-in function in Matlab.
\end{theorem}
\begin{IEEEproof}
	Please see Appendix \ref{A3}.
\end{IEEEproof}

We note that \eqref{hmft} can be further rewritten as the convolution between $f_\mathrm{r}(t)$ and $h_\mathrm{s}(t)$, which indicates that the end-to-end channel between an MF-based TX and an AP-based RX can be regarded as a linear time-invariant (LTI) system with input signal $f_\mathrm{r}(t)$, system impulse response $h_\mathrm{s}(t)$, and output signal $h_{\ss\mathrm{MF}}(t)$. We further denote $H_{\ss\mathrm{MF}}(t)$ as the expected fraction of absorbed molecules at the RX by time $t$ when vesicles are released from the center of the MF-based TX at time $t=0$ and present $H_{\ss\mathrm{MF}}(t)$ in the following corollary.
\begin{corollary}
	The expected fraction of molecules absorbed at an AP-based RX by time $t$ when vesicles are released from the center of an MF-based TX at time $t=0$ is given by
	\begin{align}\label{hmf}
	H_{\ss\mathrm{MF}}(t)&=\frac{2r_{\ss\T}r_{\ss\R}k_\mathrm{f}w_\mathrm{e}}{r_0\zeta(w_\mathrm{e})}\sum_{n=1}^{\infty}\frac{\lambda_n^3j_0(\lambda_nr_{\ss\T})}{2\lambda_nr_{\ss\T}-\sin(2\lambda_nr_{\ss\T})}\notag\\&\times\left[\xi_4(t,r_0-r_{\ss\T}-r_{\ss\R})-\xi_4(t, r_0+r_{\ss\T}-r_{\ss\R})\right],
	\end{align}
\begin{figure*}[!t]
		\begin{align}\label{xi4}
		&\xi_4(t,z)=\exp(\gamma(w_\mathrm{e})z+\zeta(w_\mathrm{e})t)\varsigma_1(t,z)-\frac{1}{2\sqrt{k_\mathrm{d}}}\exp\left(-z\sqrt{\frac{k_\mathrm{d}}{D_\sigma}}\right)\left[\left(\varpi(w_{\e})-\sqrt{k_\mathrm{d}}\right)\varsigma_2(t,z)-\left(\varpi(w_\e)+\sqrt{k_\mathrm{d}}\right)\right.\notag\\&\times\left.\left(\frac{1}{D_\mathrm{v}\lambda_n^2}\left(1-\exp(-D_\mathrm{v}\lambda_n^2t)\right)-\exp\left(2z\sqrt{\frac{k_\mathrm{d}}{D_\sigma}}\right)\varsigma_3(t,z)\right)\right]-\frac{\exp\left(-z\sqrt{\frac{k_\mathrm{d}}{D_\sigma}}\right)}{D_\mathrm{v}\lambda_n^2}\left(1-\exp\left(-D_\mathrm{v}\lambda_n^2t\right)\right). 
		\end{align}
		\hrulefill \vspace*{-1pt}
\end{figure*}

\noindent
where $\xi_4$ is given in \eqref{xi4} on the top of the next page with
 $\varsigma_2(t,z)=\int_{0}^{t}\mathrm{erf}\left(\frac{z}{\sqrt{4D_\sigma(t-u)}}-\sqrt{k_\mathrm{d}(t-u)}\right)\exp(-D_\mathrm{v}\lambda_n^2u)\mathrm{d}u$ and
	\begin{align}\label{var3}
		\varsigma_3(t,z)=&\int_{0}^{t}\exp\left(-D_\mathrm{v}\lambda_n^2u\right)\notag\\&\times\mathrm{erfc}\left(\frac{z}{\sqrt{4D_\sigma(t-u)}}+\sqrt{k_\mathrm{d}(t-u)}\right)\mathrm{d}u.
	\end{align} 
	
	We note that $\varsigma_2(t,z)$ and \eqref{var3} can be numerically calculated by the built-in function in Matlab.
\end{corollary}
\begin{IEEEproof}
	$H_{\ss\mathrm{MF}}(t)$ can be obtained from $H_{\ss\mathrm{MF}}(t)=\int_0^th_\mathrm{MF}(u)\mathrm{d}u=\int_{0}^{t}f_\mathrm{r}(u)H_\mathrm{s}(t-u)\mathrm{d}u$. By substituting \eqref{ff} and \eqref{Hst} into this expression, we obtain \eqref{hmf}.
\end{IEEEproof}

As time $t\rightarrow\infty$, the expected fraction of absorbed molecules at the RX, denoted by $H_{\ss\mathrm{MF},\infty}$, approaches a constant value. We present $H_{\ss\mathrm{MF},\infty}$ in the following corollary.
\begin{corollary}\label{c7}
	As $t\rightarrow\infty$, the  expected asymptotic fraction of absorbed molecules at an AP-based RX when vesicles are released from the center of an MF-based TX at time $t=0$ is given by 
	\begin{align}\label{hm}
	H_{\ss\mathrm{MF},\infty}\!=&\frac{r_{\ss\R}w_\mathrm{e}}{2r_{\ss\T}r_0\zeta(w_\mathrm{e})}\!\left(\!\frac{\varpi(w_\mathrm{e})}{\sqrt{k_\mathrm{d}}}\!-\!1\!\!\right)\!\!\!\left[\!\exp\!\left(\!\!-(r_0\!-\!r_{\ss\T}\!-\!r_{\ss\R})\sqrt{\!\frac{k_\mathrm{d}}{D_\sigma}}\!\right)\right.\notag\\&\left.-\exp\left(-(r_0+r_{\ss\T}-r_{\ss\R})\sqrt{\frac{k_\mathrm{d}}{D_\sigma}}\right)\right].
	\end{align}
	\begin{IEEEproof}
		Please see Appendix \ref{A2}.
	\end{IEEEproof}
\end{corollary}

\section{Analysis of Error Performance}\label{oaacp}
In this section, we consider a sequence of bits sent from the TX to the RX as described in Section \ref{bt} and derive the resulting BER. We also calculate the optimal decision threshold that minimizes the BER at the RX. We note that all results in this section are valid for both point TXs and MF-based TXs.

For simplicity, we use $H(t)$ to represent both $H_\mathrm{p}(t)$ and $H_\mathrm{MF}(t)$, where we recall that  $H_\mathrm{p}(t)$ and $H_\mathrm{MF}(t)$ represent the fraction of molecules expected to be absorbed by the RX when molecules are released from a point TX and an MF-based TX, respectively. We also use $N_{\ss\mathrm{T}}$ to represent both $N_\sigma$ and $N_\mathrm{v}\eta$, i.e., the total number of molecules released from the TX. The fraction of molecules absorbed during the $q$th bit interval for the molecules released at the beginning of the $g$th bit interval, $1\leq g\leq q$, is given by $H((q-g+1)T_\mathrm{b})-H((q-g)T_\mathrm{b})$. We denote $N_q$ as the total number of molecules absorbed during the $q$th bit interval. As we assume that the diffusion processes of all molecules are independent and all molecules have the same probability of being absorbed, $N_q$ can be approximated as a Poisson random variable (RV) when the number of emitted molecules is large and the success probability of molecules being absorbed is small \cite{le1960approximation}. Therefore, we model $N_q$ as $N_q\sim\mathrm{Poiss}(\chi)$, where $\mathrm{Poiss}(\cdot)$ denotes a Poisson distribution and $\chi$ is given by 
\begin{align}
	\chi=N_{\ss\T}\sum_{g=1}^{q}b_g\left(H((q-g+1)T_\mathrm{b})-H((q-g)T_\mathrm{b})\right).
\end{align}

At the RX, we adopt a threshold detector that compares $N_q$ with a decision threshold for demodulation of $b_q$. We model the threshold detector as follows
\begin{align}\label{bk}
\hat{b}_q=\left\{\begin{array}{lr}
1, ~~\mathrm{if}~N_q\geq\Psi,
\\
0, ~~\mathrm{if}~N_q<\Psi,
\end{array}
\right.
\end{align}
where $\hat{b}_q$ is the demodulated bit for $b_q$ and $\Psi$ is the decision threshold. We then denote $\Phi[q|\mathbf{b}_{1:q-1}]$ as the BER of the $q$th bit given the previously transmitted sequence $\mathbf{b}_{1:q-1}$. According to \eqref{bk}, $\Phi[q|\mathbf{b}_{1:q-1}]$ is given by
\begin{align}\label{phi}
\Phi[q|\mathbf{b}_{1:q-1}]&=P_1\mathrm{Pr}\left(N_q<\Psi|b_q=1, \mathbf{b}_{1:q-1}\right)\notag\\&+P_0\mathrm{Pr}\left(N_q\geq\Psi|b_q=0, \mathbf{b}_{1:q-1}\right).
\end{align}

We denote $\Psi^*[q|\mathbf{b}_{1:q-1}]$ as the optimal decision threshold that minimizes $\Phi[q|\mathbf{b}_{1:q-1}]$. According to \cite[Eq. (25)]{ahmadzadeh2015analysis}, we obtain $\Psi^*[q|\mathbf{b}_{1:q-1}]$ as
\begin{align}
	\Psi^*[q|\mathbf{b}_{1:q-1}]=\left\lceil\frac{\ln\frac{P_0}{P_1}+N_\sigma H(\phi)}{\ln\frac{\chi_1|\mathbf{b}_{1:q-1}}{\chi_0|\mathbf{b}_{1:q-1}}}\right\rceil,
\end{align}
where $\chi_1|\mathbf{b}_{1:q-1}$ and $\chi_0|\mathbf{b}_{1:q-1}$ are the expected total number of molecules absorbed within the $q$th bit interval assuming that $b_q=1$ and $b_q=0$ was transmitted, respectively. They are given by
\begin{align}
\chi_1|\mathbf{b}_{1:q-1}=&N_{\ss\T}\sum_{g=1}^{q-1}b_g\left(H((q-g+1)T_\mathrm{b})-H((q-g)T_\mathrm{b})\right)\notag\\&+N_\sigma H(T_\mathrm{b})
\end{align}
and
\begin{align}
\chi_0|\mathbf{b}_{1:q-1}\!=\!N_{\ss\T}\!\sum_{g=1}^{q-1}\!b_g\!\left(\!H(\!(q\!-\!g\!+\!1)T_\mathrm{b})-H((q-g)T_\mathrm{b})\right),
\end{align}
where $\lceil\cdot\rceil$ represents the ceiling function \cite{isgur2012nested}.

We further consider the average optimal decision threshold, denoted by $\overline{\Psi}^*$, that minimizes the average BER over all realizations of $\mathbf{b}_{1:q}$ and all bits $b_q, q\in\{1,...,Q\}$. Thereby, we have
\begin{align}\label{psi}
	\overline{\Psi}^*=\mathop{\min}_{\Psi}\frac{1}{Q}\sum_{q=1}^{Q}\frac{\sum_{\mathbf{b}_{1:q-1}}\Phi[q|\mathbf{b}_{1:q-1}]}{2^{q-1}}.
\end{align}
We note that the optimization of $\overline{\Psi}^*$ in \eqref{psi} can be obtained numerically by a simple one-dimensional search.

\section{Numerical Results}\label{nr}

\begin{table}[!t]
	\newcommand{\tabincell}[2]{\begin{tabular}{@{}#1@{}}#2\end{tabular}}
	\centering
	\caption{Simulation Parameters for Numerical Results}\label{tabl}
	\scalebox{0.94}{\begin{tabular}{|c|c|c|c|}
			\hline
			\textbf{Parameter}&\textbf{Variable}&\textbf{Value}&\textbf{Reference}\\
			\hline
			\tabincell{c}{Ratio of the total area \\of APs to the RX surface}&$\mathcal{A}$&0.05, 0.1, 0.15&\cite{lindsay2017first}\\
			\hline
			Number of APs&$N_\mathrm{p}$&1$\sim$1000&\cite{li2004cellular}\\
			\hline
			Radius of the RX&$r_{\ss\R}$&$10\;\mu\mathrm{m}$&\cite{hat2011b}\\
			\hline
			\tabincell{c}{Distance between the center of \\RX and point TX/center of \\MF-based TX}&$r_0$&$20\;\mu\mathrm{m}$&\cite{yilmaz2014three}\\
			\hline
			\tabincell{c}{Number of emitted molecules \\ from the point TX for bit 1}&$N_\sigma$&1000& \cite{heren2015effect}\\
			\hline
			Molecule degradation rate&$k_\mathrm{d}$&$0.8\;\mathrm{s}^{-1}$&\cite{deng2015modeling}\\
			\hline
			Molecule diffusion coefficient& $D_\sigma$&$79.4\;\mu\mathrm{m}^2/\mathrm{s}$&\cite{yilmaz2014three}\\
			\hline
			Radius of the MF-based TX&$r_{\ss\T}$&$5\;\mu\mathrm{m}$&\cite{hat2011b}\\
			\hline
			\tabincell{c}{Number of released vesicles\\for bit 1}&$N_\mathrm{v}$&200&\cite{huang2021fusion}\\
			\hline
			\tabincell{c}{Number of molecules stored \\ in each vesicle}&$\eta$&5&\cite{huang2021fusion}\\
			\hline
			Diffusion coefficient of vesicles&$D_\mathrm{v}$&$9\;\mu\mathrm{m}^2/\mathrm{s}$&\cite{kyoung2008vesicle}\\
			\hline
			\tabincell{c}{Forward reaction rate of \\ vesicles and TX membrane}&$k_\mathrm{f}$&$30\;\mu\mathrm{m}/\mathrm{s}$&\cite{huang2021fusion}\\
			\hline
			Length of binary bit sequence&$Q$&10&\\
			\hline
			Bit interval length&$T_\mathrm{b}$&$0.8\;\mathrm{s}$&\\
			\hline
			Probability of transmitting 0/1&$P_0$/$P_1$&0.5&\\
			\hline

		\end{tabular}
	}
\end{table}

In this section, we present numerical results to validate our theoretical analysis and to offer insights for MC system design. Specifically, we use PBSs to simulate the random diffusion of molecules. In our simulations, if the position of a molecule at the end of the current simulation step is inside the RX volume, we assume that this molecule has hit the RX surface in this simulation step. The coordinates of the hitting points on the RX surface are calculated by using \cite[Eqs. (36)-(38)]{huang2021fusion}. If the coordinates of the hitting point of a molecule are inside an AP, we treat this molecule as an absorbed molecule. Otherwise, the molecule is reflected back to the position it was at the start of the current simulation step \cite{ahmadzadeh2016comprehensive}. To model the location of the point TX (or the center of the MF-based TX), we fix the distance between the point TX (or the center of the MF-based TX) and the RX to $r_0$ and randomly generate the location of the center of the TX for each realization. The simulation framework for modeling the molecule release for the MF-based TX is detailed in \cite[Sec. VI]{huang2021fusion}. We choose the simulation time step as $\Delta t=10^{-7}\;\mathrm{s}$ and all results are averaged over 1000 realizations. Throughout this section, we set the simulation parameters as shown in Table \ref{tabl}, unless otherwise stated. We specify the value of $N_\mathrm{p}$ and the locations of the APs in each figure. From Figs. \ref{term}, \ref{frac2}, and \ref{MF}, we observe that the simulation results (denoted by markers) match well with the derived analytical curves (denoted by solid and dashed lines) generated based on the results presented in Sections \ref{cirp} and \ref{ac}, which validates the accuracy of our CIR analysis. In all figures, we apply \eqref{G1} to generate the analytical curves if the APs have identical sizes, \eqref{Gc} if the APs have different sizes, and \eqref{g1a} if there is a single AP, unless otherwise stated.

\subsection{Point TX}\label{nrpt}

\begin{figure}[!t]
	\begin{center}
		\includegraphics[height=2.2in,width=0.86\columnwidth]{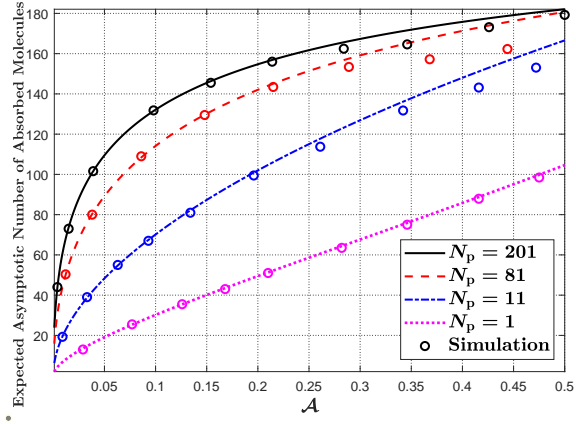}
		\caption{Expected asymptotic number of absorbed molecules versus $\mathcal{A}$ for different values of $N_\mathrm{p}$, where APs have identical sizes and are evenly distributed.}\label{ga}\vspace{-0.5em}
	\end{center}
	\vspace{-4mm}
\end{figure}

In this subsection, we focus on the point TX model. We first investigate the range of $\mathcal{A}$ for which our analytical expressions are accurate. In Fig. \ref{ga}, we plot $N_\sigma H_{\mathrm{p},\infty}$ versus $\mathcal{A}$ for different values of $N_\mathrm{p}$. First, when $N_\mathrm{p}=\{11, 81, 201\}$, we observe that the simulation results match well with the analytical curves for $\mathcal{A}\leq0.2$. As $\mathcal{A}$ increases, a gap between the analytical curves and simulation results emerges. This is because the range of $\mathcal{A}$, where the CIR analysis is accurate, is limited since we ignore the term $\mathcal{O}\left(\kappa^2\ln\left(\frac{\kappa}{2}\right)\right)$ in \eqref{Gc} when calculating $G_\mathrm{p}$. We note that the range of $\mathcal{A}$ for which our analytical results are accurate aligns with the typical coverage of cell membranes with receptors. For example, the ratio of the total area of a liver cell covered by insulin receptors to the total cell surface is around 0.0026 \cite{lauffenburger1996receptors,parang2001mechanism}, which is well within the range of $\mathcal{A}$ where our analytical expressions are valid. Second, we observe that the simulation results always match well with the analytical curve when $N_\mathrm{p}=1$. This is because $\mathcal{O}(\kappa^3\ln\kappa)$ in \eqref{g1a} has a higher order than $\mathcal{O}\left(\kappa^2\ln\left(\frac{\kappa}{2}\right)\right)$ in \eqref{Gc} such that \eqref{g1a} is accurate even when the AP has a large size. 

\begin{figure}[!t]
	\centering	
	\subfigure[Expected molecule hitting rate]{
		\begin{minipage}[t]{1\linewidth}
			\centering
			\includegraphics[width=3in]{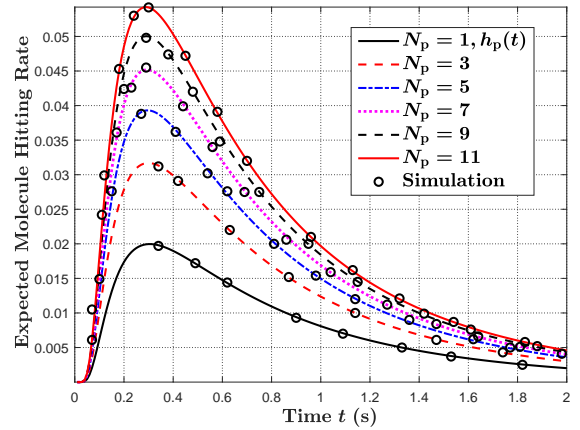}
			\label{a}
		\end{minipage}%
	}%
	\quad
	\subfigure[Expected number of absorbed molecules]{
		\begin{minipage}[t]{1\linewidth}
			\centering
			\includegraphics[width=3in]{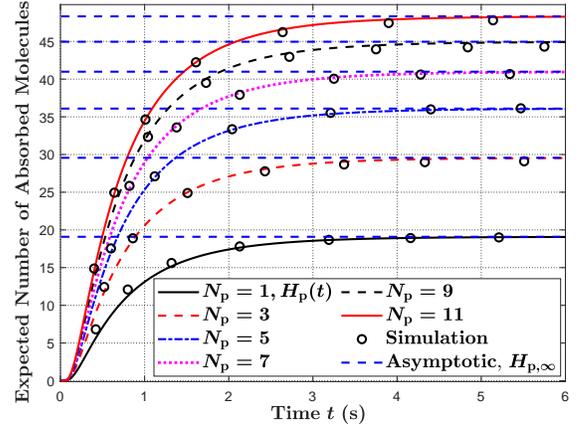}
			\label{b}
		\end{minipage}%
	}
	\centering
	\caption{Expected molecule hitting rate at time $t$ and expected number of molecules absorbed until time $t$ versus time $t$ for different values of $N_\mathrm{p}$ and $\mathcal{A}=0.05$.}\label{term}
\end{figure}

 In Fig. \ref{a}, we show the expected molecule hitting rate at the RX at time $t$, and in Fig. \ref{b}, we depict the expected number of molecules absorbed at the RX by time $t$. We set $\mathcal{A}=0.05$ and increase the number of APs from 1 to 11. All APs have the same size and are evenly distributed on the RX surface. We observe that the expected hitting rate in Fig. \ref{a} and the expected number of absorbed molecules in Fig. \ref{b} increase with $N_\mathrm{p}$. This behaviour is expected since a larger number of APs on the RX surface can absorb more molecules.
 
 \begin{figure}[!t]
 	\begin{center}
 		\includegraphics[height=2.2in,width=0.86\columnwidth]{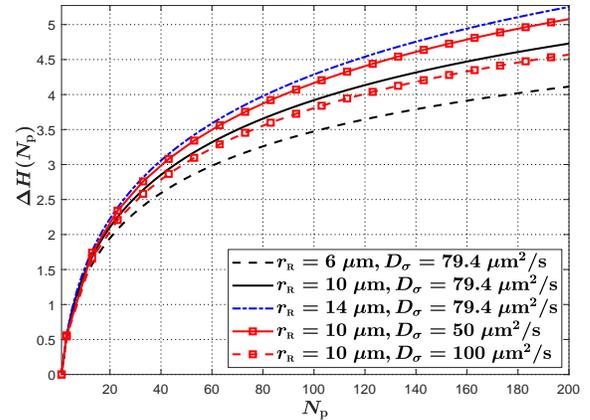}
 		\caption{$\Delta H(N_\mathrm{p})$ versus $N_\mathrm{p}$ for varying $r_{\ss\R}$ and $D_\sigma$, where $\mathcal{A}=0.05$, identical APs are evenly distributed on the RX surface.}\label{increment}\vspace{-0.5em}
 	\end{center}
 	\vspace{-4mm}
 \end{figure}
 
 To further investigate the observation that the number of absorbed molecules increases with increasing $N_\mathrm{p}$, we define the relative increase in the number of absorbed molecules as $N_\mathrm{p}$ increases, denoted by $\Delta H(N_\mathrm{p})$, as follows
 \begin{align}\label{in}
 \Delta H(N_\mathrm{p})&=\frac{H_{\mathrm{p},\infty}-\tilde{H}_{\mathrm{p},\infty}}{\tilde{H}_{\mathrm{p},\infty}}\!\notag\\&=\frac{w_{\mathrm{e},\ss N_\mathrm{p}}\zeta(w_{\mathrm{e},1})\left(\gamma(w_{\mathrm{e},\ss N_\mathrm{p}})-\sqrt{\frac{k_\mathrm{d}}{D_\sigma}}\right)}{w_{\mathrm{e},1}\zeta(w_{\mathrm{e},\ss N_\mathrm{p}})\left(\gamma(w_{\mathrm{e},1})-\sqrt{\frac{k_\mathrm{d}}{D_\sigma}}\right)}-1,
 \end{align}
 where $\tilde{H}_{\mathrm{p},\infty}=H_{\mathrm{p},\infty}|_{N_\mathrm{p}=1}$ represents the expected asymptotic number of absorbed molecules when $N_\mathrm{p}=1$, and $w_{\mathrm{e},1}$ and $w_{\mathrm{e},\ss N_\mathrm{p}}$ represent the effective reaction rates of RXs with a single AP and with $N_\mathrm{p}$ APs, respectively. In Fig. \ref{increment}, we plot $\Delta H(N_\mathrm{p})$ versus $N_\mathrm{p}$ for different $r_{\ss\R}$ and $D_\sigma$ and note that larger $\Delta H(N_\mathrm{p})$ imply a larger relative increase in the number of absorbed molecules. First, we observe that $\Delta H(N_\mathrm{p})$ increases with increasing $N_\mathrm{p}$. This is because larger $N_\mathrm{p}$ result in more absorbed molecules. We also observe that the slope of $\Delta H(N_\mathrm{p})$ becomes smaller as $N_\mathrm{p}$ becomes larger, which indicates that the increase in the number of absorbed molecules becomes slower for larger $N_\mathrm{p}$. Second, we observe that $\Delta H(N_\mathrm{p})$ increases if $r_{\ss\R}$ increases and $N_\mathrm{p}$ is fixed. This is because a larger RX leads to a larger AP area such that the relative increase in the number of absorbed molecules is larger. Third, we observe that $\Delta H(N_\mathrm{p})$ increases if $D_\sigma$ decreases. This is because larger $N_\mathrm{p}$ give slowly diffusing molecules more chances to be absorbed than fast diffusing molecules, such that the relative increase in the number of absorbed molecules for smaller $D_\sigma$ is larger.

\begin{figure}[!t]
	\begin{center}
		\includegraphics[height=2.2in,width=0.86\columnwidth]{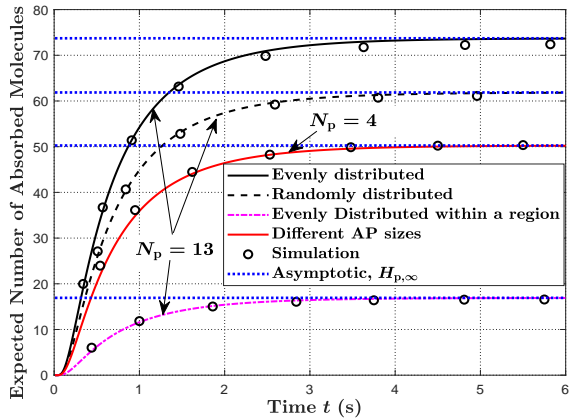}
		\caption{Expected number of molecules absorbed by the RX until time $t$ versus time $t$ for different distributions and sizes of APs, where $\mathcal{A}=0.1$.}\label{frac2}\vspace{-0.5em}
	\end{center}
	\vspace{-4mm}
\end{figure} 

In Fig. \ref{frac2}, we evaluate the impact of different AP sizes and distributions on the RX surface. First, we define $\mathcal{A}_i$ as the ratio of the area of $\mathrm{AP}_i$ to the area of the RX surface and set $N_\mathrm{p}=4$, $\mathcal{A}_1=0.01$, $\mathcal{A}_2=0.02$, $\mathcal{A}_3=0.03$, and $\mathcal{A}_4=0.04$ to validate our analytical expressions when the APs have different sizes. Then, we set the locations of the four APs to $\vec{l}_1=[10\;\mu\mathrm{m}, \pi/2, \pi]$, $\vec{l}_2=[10\;\mu\mathrm{m}, \pi/2, \pi/2]$, $\vec{l}_3=[10\;\mu\mathrm{m}, \pi/2, 0]$, and $\vec{l}_4=[10\;\mu\mathrm{m}, \pi/2, 3\pi/2]$. We observe from Fig. \ref{frac2} that the simulation results match well with the analytical curves, which demonstrates the accuracy of \eqref{com} for APs of different sizes. Second, we set $\mathcal{A}=0.1$ and $N_\mathrm{p}=13$ for different AP distributions and assume that the APs have identical size. Here, we consider three types of AP distributions: i) APs evenly distributed across the entire RX surface, ii) APs randomly distributed across the entire RX surface, and iii) APs evenly distributed within a region of the RX surface.
In this paper, we apply the Fibonacci lattice \cite{gonzalez2010measurement} to determine the locations of the evenly distributed APs across the entire RX surface. Specifically, the location of $\mathrm{AP}_i$ is given by
\begin{align}
\theta_i\!=\!\frac{\pi}{2}\!-\!\arcsin\!\left(\!\frac{2(i\!-\!\mathcal{B}\!-1)}{N_\mathrm{p}}\!\right), \varphi_i\!=\!\frac{4\pi(i\!-\!\mathcal{B}\!-\!1)}{1+\sqrt{5}},
\end{align}
where integer $\mathcal{B}$ is given by $\mathcal{B}=(N_\mathrm{p}-1)/2$. The locations of evenly distributed APs across the entire RX surface can also be generated by the built-in function \textit{SpherePoints} in Mathematica. For the APs evenly distributed within a region of the RX surface, we assume that the APs are distributed within the area defined by $\theta_i\in[2.812, 3.471]$ and $\varphi_i\in[0, 2\pi]$. From Fig. \ref{frac2}, we observe that APs evenly distributed across the entire RX surface yield a larger expected number of absorbed molecules compared to the other two distributions. We next explain this observation from two perspectives:
\begin{figure}[!t]
	\centering	
	\subfigure[APs have the identical size]{
		\begin{minipage}[t]{1\linewidth}
			\centering
			\includegraphics[height=1.8in,width=0.65\columnwidth]{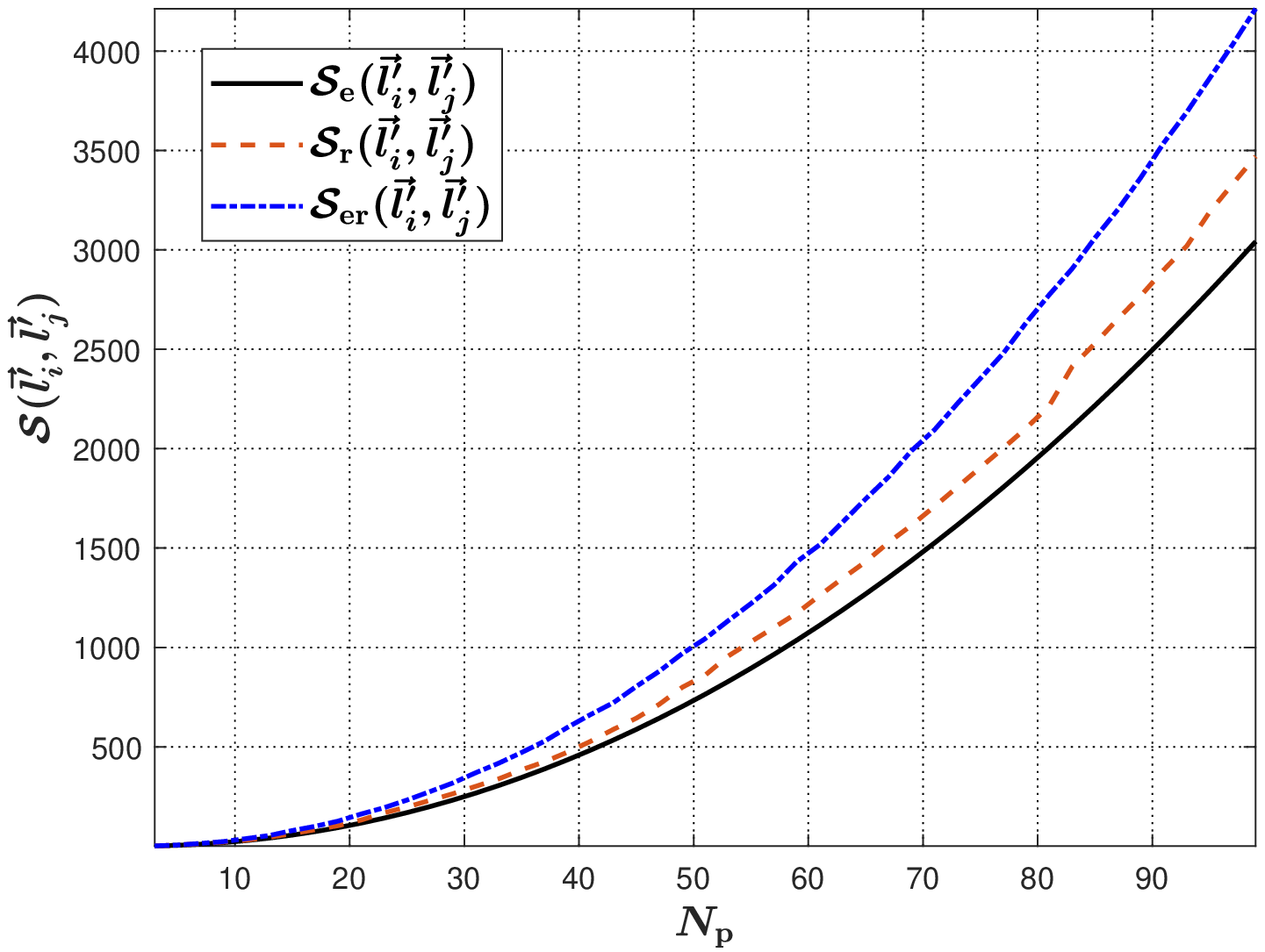}
			\label{fc}
		\end{minipage}%
	}%
	\quad
	\subfigure[APs have different sizes]{
		\begin{minipage}[t]{1\linewidth}
			\centering
			\includegraphics[height=2.5in,width=1\columnwidth]{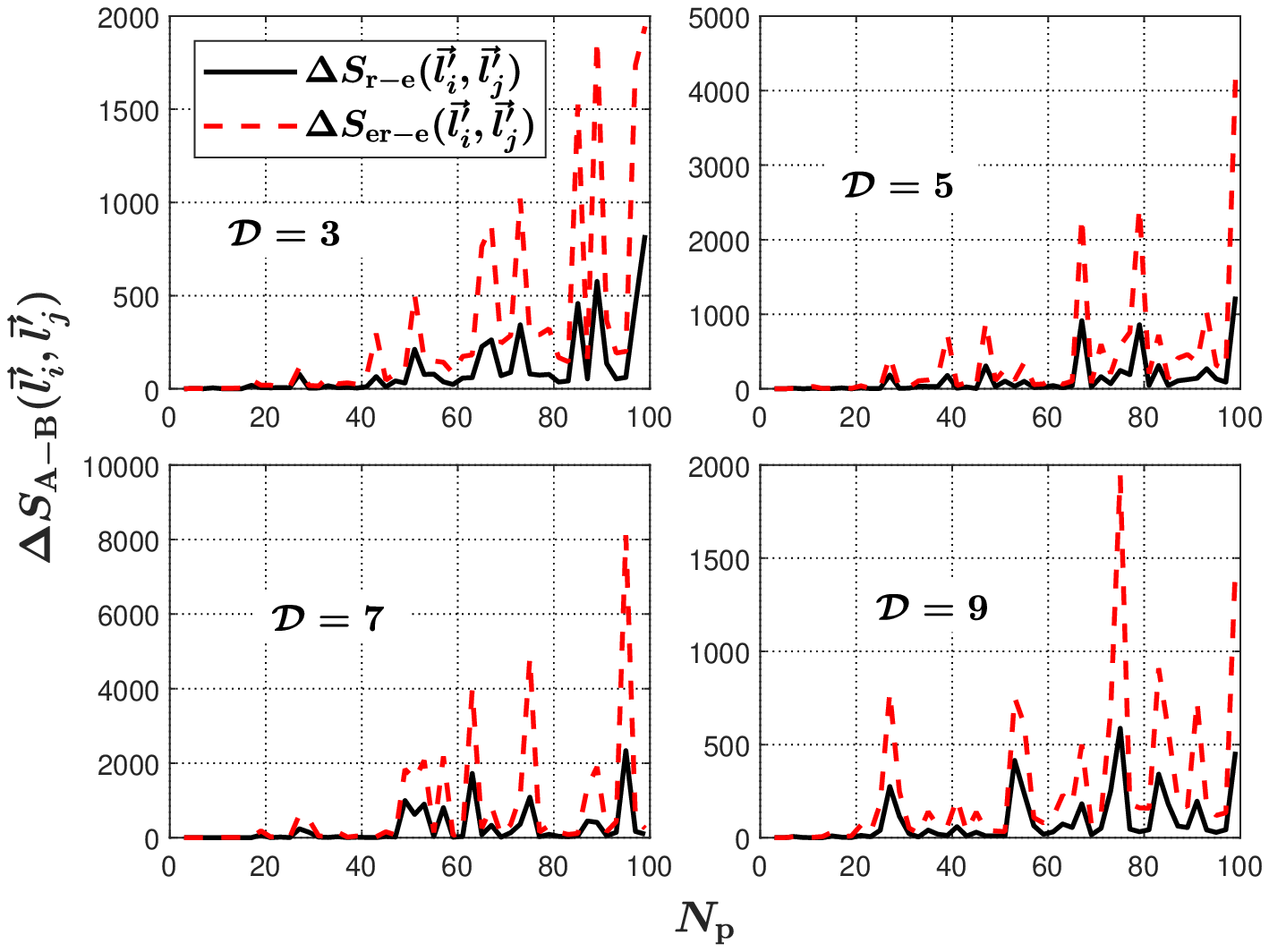}
			\label{fc2}
		\end{minipage}%
	}
	\centering
	\caption{(a): $\mathcal{S}(\vec{l}_i', \vec{l}_j')$ versus $N_\mathrm{p}$ for different AP distributions on the RX surface. (b): $\Delta\mathcal{S}_{\mathrm{A}-\mathrm{B}}(\vec{l}_i', \vec{l}_j')$ versus $N_\mathrm{p}$. We set $\mathcal{A}=0.1$.}\label{fcom}
\end{figure}
\begin{itemize}
	\item Theoretical perspective: According to \eqref{com}, the expected number of absorbed molecules is directly determined by $w_\mathrm{e}$. In particular, a larger value of $w_\mathrm{e}$ results in a larger number of molecules expected to be absorbed by the RX. Moreover, based on \eqref{wee}, we observe that $w_\mathrm{e}$ is a monotonically increasing function with respect to $G_\mathrm{p}$. Furthermore, we set $\mathcal{S}(\vec{l}_i', \vec{l}_j')=\sum_{i=1}^{N_\mathrm{p}}\sum_{j=i+1}^{N_\mathrm{p}}m_im_j\mathcal{F}(\vec{l}_i', \vec{l}_j')$, where $\mathcal{F}(\vec{l}_i', \vec{l}_j')$ is given in \eqref{F}. In \eqref{Gc}, we observe that only $\mathcal{S}(\vec{l}_i', \vec{l}_j')$ is related to the distributions of the APs, and a smaller $\mathcal{S}(\vec{l}_i', \vec{l}_j')$ leads to a larger $G_\mathrm{p}$. Therefore, the distribution of APs that results in a smaller $\mathcal{S}(\vec{l}_i', \vec{l}_j')$ is expected to lead to a larger number of absorbed molecules at the RX. To compare $\mathcal{S}(\vec{l}_i', \vec{l}_j')$ for the three considered distributions, we write $\mathcal{S}(\vec{l}_i', \vec{l}_j')$ as $\mathcal{S}_\mathrm{e}(\vec{l}_i', \vec{l}_j')$ and $\mathcal{S}_\mathrm{r}(\vec{l}_i', \vec{l}_j')$ if the APs are evenly and randomly distributed across the entire RX surface, respectively, and write $\mathcal{S}(\vec{l}_i', \vec{l}_j')$ as $\mathcal{S}_\mathrm{er}(\vec{l}_i', \vec{l}_j')$ if the APs are evenly distributed within a region of the RX surface. Moreover, we define $\Delta\mathcal{S}_{\mathrm{A}-\mathrm{B}}(\vec{l}_i', \vec{l}_j')=\mathcal{S}_\mathrm{A}(\vec{l}_i', \vec{l}_j')-\mathcal{S}_\mathrm{B}(\vec{l}_i', \vec{l}_j')$, $A, B\in\{\mathrm{r}, \mathrm{e}, \mathrm{er}\}$, to evaluate the difference between the $\mathcal{S}(\vec{l}_i', \vec{l}_j')$ for two different distributions. In Fig. \ref{fc}, we plot $\mathcal{S}_\mathrm{e}(\vec{l}_i', \vec{l}_j')$, $\mathcal{S}_\mathrm{r}(\vec{l}_i', \vec{l}_j')$, and $\mathcal{S}_\mathrm{er}(\vec{l}_i', \vec{l}_j')$ versus $N_\mathrm{p}$ for APs of identical size, and in Fig. \ref{fc2}, we show $\Delta\mathcal{S}_{\mathrm{A}-\mathrm{B}}(\vec{l}_i', \vec{l}_j')$ versus $N_\mathrm{p}$ for APs of different sizes.  For APs evenly distributed within a region, we set the size of the area of this region to $40\%$ of the entire area of the RX surface. For APs of different sizes, we randomly generate $N_\mathrm{p}$ integers from 1 to $\mathcal{D}$ and denote $\mathcal{G}_i$ as the $i$th integer. We calculate $\mathcal{A}_i$ as $\mathcal{A}_i=\mathcal{A}\mathcal{G}_i/\sum_{i=1}^{N_\mathrm{p}}\mathcal{G}_i$. In Fig. \ref{fc}, we observe that $\mathcal{S}_\mathrm{e}(\vec{l}_i', \vec{l}_j')$ is always smaller than $\mathcal{S}_\mathrm{r}(\vec{l}_i', \vec{l}_j')$ and $\mathcal{S}_\mathrm{er}(\vec{l}_i', \vec{l}_j')$. In Fig. \ref{fc2}, we apply different values of $\mathcal{D}$ and observe that $\mathcal{S}_{\mathrm{r}-\mathrm{e}}(\vec{l}_i', \vec{l}_j')>0$ and  $\mathcal{S}_{\mathrm{er}-\mathrm{e}}(\vec{l}_i', \vec{l}_j')>0$ for all $N_\mathrm{p}$ and $\mathcal{D}$. These observations imply that, for evenly distributed APs, $\mathcal{S}(\vec{l}_i', \vec{l}_j')$ is always smaller than for randomly distributed APs and APs evenly distributed within a region, regardless of whether the APs have identical size or different sizes. Therefore, evenly distributed APs result in a larger number of absorbed molecules than the other two distributions.
	\item Physical perspective: Evenly distributed APs ensure that the APs are equally spaced covering the entire RX surface. Thus, for most locations of the TX, there are APs located in the region closest to the TX such that the probability that a molecule hits an AP is maximized. When the APs are randomly distributed or evenly distributed within a region, only those TX locations that happen to be close to a region on the RX surface with APs have a high probability of molecules hitting an AP, while other TX locations have a low probability. Therefore, the expected probability of molecules hitting an AP for randomly distributed APs and APs distributed within a region is lower than that for evenly distributed APs.
\end{itemize}

\begin{figure}[!t]
	\begin{center}
		\includegraphics[height=2.2in,width=0.86\columnwidth]{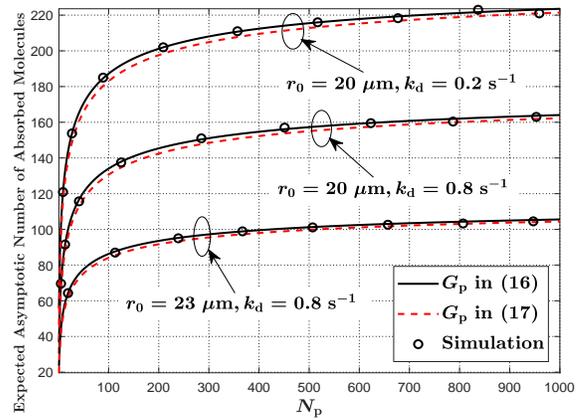}
		\caption{Expected asymptotic number of absorbed molecules versus $N_\mathrm{p}$ for different values of $r_0$ and $k_\mathrm{d}$, where $\mathcal{A}=0.15$ and APs of identical size are evenly distributed on the RX surface.}\label{df}\vspace{-0.5em}
	\end{center}
	\vspace{-4mm}
\end{figure}

We recall that when the APs have the same size and are evenly distributed on the RX surface, $G_\mathrm{p}$ in \eqref{G1} can be further simplified to \eqref{Gnl}. To verify \eqref{Gnl}, in Fig. \ref{df}, we plot $N_\sigma H_{\mathrm{p},\infty}$, i.e., the expected asymptotic number of absorbed molecules, versus $N_\mathrm{p}$ for different values of $r_0$ and $k_\mathrm{d}$, where $G_\mathrm{p}$ in $H_{\mathrm{p},\infty}$ is given by \eqref{G1} and \eqref{Gnl}, respectively. We observe that the gap between $H_{\mathrm{p},\infty}$ calculated based on \eqref{G1} and \eqref{Gnl} is extremely small, which confirms the accuracy of \eqref{Gnl}. 

\subsection{MF-based TX}

\begin{figure}[!t]
	\centering
	
	\subfigure[Expected molecule hitting rate]{
		\begin{minipage}[t]{1\linewidth}
			\centering
			\includegraphics[width=3in]{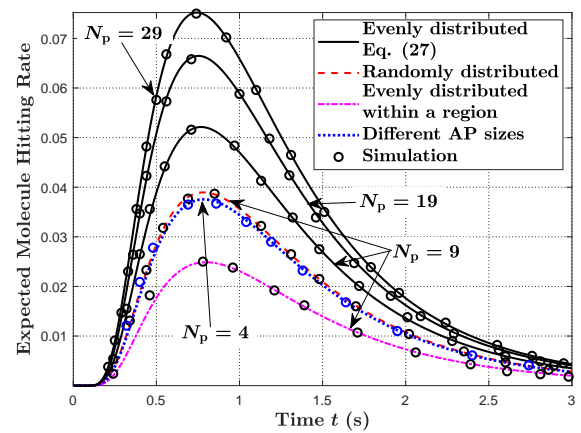}
			\label{rM}
		\end{minipage}%
	}%
\quad 
	\subfigure[Expected number of absorbed molecules]{
		\begin{minipage}[t]{1\linewidth}
			\centering
			\includegraphics[width=3in]{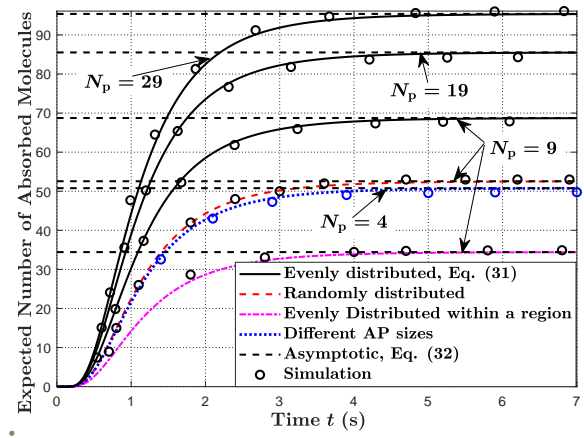}
			\label{fM}
		\end{minipage}%
	}
	\centering
	\caption{Expected molecule hitting rate at time $t$ and expected number of molecules absorbed by the RX until time $t$ versus time $t$ for different numbers, locations, and sizes of APs, where $\mathcal{A}=0.1$.}\label{MF}
\end{figure}

In this subsection, we consider the MF-based TX. In Fig. \ref{rM}, we verify \eqref{hmm}, \eqref{hmf}, and \eqref{hm} by depicting the expected molecule hitting rate versus time $t$, and in Fig. \ref{fM}, we show the expected number of absorbed molecules versus time $t$ for different numbers, locations, and sizes of APs. We use the same AP locations and sizes as in Fig. \ref{frac2}, considering APs evenly distributed within a region and APs having different sizes. From Fig. \ref{MF}, we observe that the expected molecule hitting rate and the expected number of absorbed molecules increase with increasing $N_\mathrm{p}$. We also observe that evenly distributed APs achieve a larger molecule hitting rate and a larger number of absorbed molecules than randomly distributed APs and APs evenly distributed within a region. These observations are consistent with the observations for point TXs. 

\subsection{Comparison of Different TX-RX Models}
In this subsection, we compare three different combinations of TXs and RXs, namely i) a point TX and a fully absorbing RX (referred to as PTFR), ii) a point TX and an AP-based RX (referred to as PTAR), and iii) an MF-based TX and an AP-based RX (referred to as MTAR). We compare these three combinations with respect to the CIRs and average BERs.

\begin{figure}[!t]
	\begin{center}
		\includegraphics[height=2.2in,width=0.86\columnwidth]{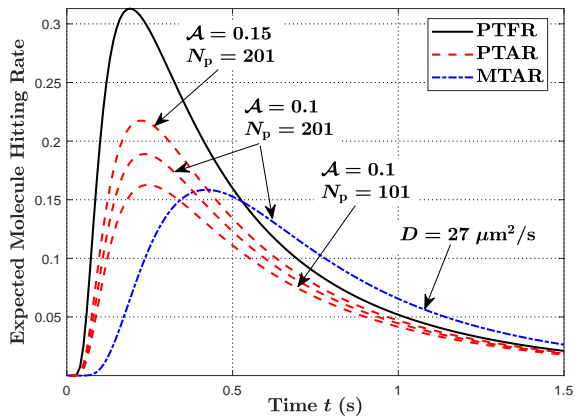}
		\caption{Expected molecule hitting rate versus time $t$ for three different TX-RX models, where identical APs are evenly distributed on the RX surface.}\label{ht}\vspace{-0.5em}
	\end{center}
	\vspace{-4mm}
\end{figure} 

In Fig. \ref{ht}, we plot the expected molecule hitting rate versus time $t$ for the three considered combinations of TXs and RXs, where the expected molecule hitting rate of PTFR is given by \cite[Eq. (9)]{heren2015effect}. First, by comparing the molecule hitting rates of PTFR and PTAR, we observe that the peak value of the molecule hitting rate decreases and the tail becomes shorter when a fully absorbing RX is replaced by an AP-based RX. This is due to the fact that the limited coverage and the finite number of APs on the RX surface reduce the number of molecules absorbed by the RX within a time period. We also observe that, for the PTAR model, the peak value of the molecule hitting rate becomes larger and the tail is longer when $\mathcal{A}$ or the number of APs increases. Second, by comparing the molecule hitting rates of PTFR and MTAR, we observe that the peak value of the molecule hitting rate of MTAR is smaller and the tail is longer. This is because the release of molecules from the MF-based TX is slower compared to the point TX, and the absorption of molecules at the AP-based RX is limited compared to the fully absorbing RX. In summary, compared with the ideal PTFR model, the practical MFAR model results in a smaller amplitude and a longer duration of the molecule hitting rate curve.

\begin{figure}[!t]
	\centering
	
	\subfigure[Average BER versus $\Psi$]{
		\begin{minipage}[t]{1\linewidth}
			\centering
			\includegraphics[width=3in]{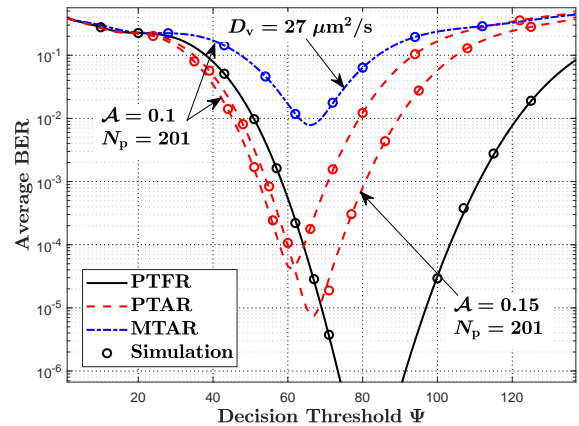}
			\label{berf}
		\end{minipage}%
	}%
\quad
	\subfigure[Average BER versus $T_\mathrm{b}$]{
		\begin{minipage}[t]{1\linewidth}
			\centering
			\includegraphics[width=3in]{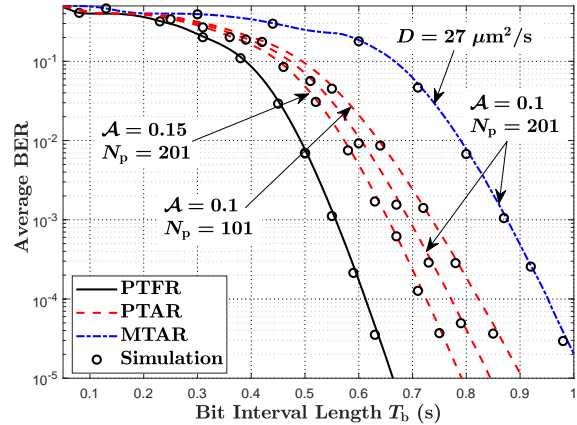}
			\label{interval}
		\end{minipage}%
	}
	\centering
	\caption{(a): Average BER versus decision threshold $\Psi$ for three TX-RX models. (b): Average BER versus bit interval length $T_\mathrm{b}$ for three TX-RX models, where information is demodulated at the average optimal decision threshold $\overline{\Psi}^*$. Identical APs are evenly distributed on the RX surface.}\label{aber}
\end{figure}

We show the average BER versus the decision threshold in Fig. \ref{berf} and versus the bit interval length in Fig. \ref{interval} for the three considered TX-RX models. Due to the long PBS times, for this figure, we employ Monte Carlo simulation where the Poisson approximation is used to generate the number of absorbed molecules. We note that the derived analytical result for the CIR has been verified by PBSs and the Poisson approximation for the number of absorbed molecules has been widely adopted in previous studies \cite{jamali2019channel}. The average BER for PTFR is plotted by replacing $H(t)$ in Section \ref{oaacp} with \cite[Eq. (12)]{heren2015effect}. In Fig. \ref{interval}, detection is performed based on $\overline{\Psi}^*$ obtained in \eqref{psi}. First, in Fig. \ref{berf}, by comparing the average BERs for PTFR and PTAR, we observe that the BER for PTAR is slightly lower when $\Psi$ is small and becomes much higher when $\Psi$ is large. This is because the number of absorbed molecules is lower and the intersymbol interference (ISI) is smaller for AP-based RXs. Hence, the average BER is lower when the decision threshold is small. Second, in Fig. \ref{interval}, we observe that the average BER of PTFR is always lower than that of PTAR when detection is performed based on the optimal decision threshold. Third, in both Fig. \ref{berf} and Fig. \ref{interval}, we observe that the average BER of MTAR is much larger than that of PTFR. This is because the release of molecules from the MF-based TX is slower compared to the point TX, and the absorption of molecules at the AP-based RX is limited compared to the fully absorbing RX. The second and third observations imply that the widely-investigated BER analysis of the ideal PTFR model is not accurate when we consider a more practical combined TX-RX model.

\section{Conclusion}\label{con}
In this paper, we studied RXs that are covered by multiple heterogeneous APs with different sizes and arbitrary locations. We also considered two TXs models, namely point TXs and MF-based TXs. We obtained closed-form expressions for the expected CIR in the presence of molecule degradation. Our simulation results confirmed the accuracy of the derived CIR expressions. Using numerical results, we showed that when the ratio of the total area of the APs to the area of the RX surface is fixed, the expected number of absorbed molecules increases with the number of APs. Moreover, we compared three spatial AP distributions and showed that evenly distributed APs result in a larger number of absorbed molecules than the other two considered distributions. In addition, we compared three combinations of TXs and RXs. Our results showed that compared to the ideal model consisting of a point TX and fully absorbing RX, the practical model with MF-based TX and AP-based RX results in a much lower CIR and a much higher average BER. This underscores the limitations of the ideal TX-RX model for application in practical MC systems. 

Interesting topics for future work include i) optimizing the spatial distributions and sizes of the APs on the RX surface to maximize the number of absorbed molecules, and ii) considering the occupancy of heterogeneous receptors by molecules and investigating its impact on the received signal. Additionally, as the typical cell membrane is composed of a mixture of phospholipids in a fluid phase \cite{singer1972fluid}, future work can consider the movement of receptors on the RX surface and incorporate this phenomenon into MC system design.

\appendices
\section{Proof of Corollary \ref{asy}}\label{AA}
According to \eqref{Hu}, we have 
\begin{align}\label{H}
H_\mathrm{u,\infty}(w)=\frac{r_{\ss\mathrm{R}}w}{r_0}
\left[\alpha_1(\infty)-\alpha_2(\infty,w)\right].
\end{align}
Based on \eqref{d1}, we have $\alpha_1(\infty)=\exp(-\beta)/\sqrt{k_\mathrm{d}D_\sigma}$. To obtain $\alpha_2(\infty,w)$, we need to determine $\psi_1(\infty,w)$ and $\psi_2(\infty,w)$ in \eqref{d2}. In \eqref{e1}, if $\zeta(w)\leq0$, we have $\psi_1(\infty,w)=0$. If $\zeta(w)>0$, we obtain
\begin{align}\label{e1a}
&\psi_1(\infty,w)=2\gamma(w)\exp(\gamma(w)(r_0-r_{\ss\mathrm{R}}))\notag\\&\times\lim_{t\rightarrow\infty}\frac{\mathrm{erfc}\left(\frac{\varepsilon}{\sqrt{t}}+\gamma(w)\sqrt{D_\sigma t}\right)}{\exp(-\zeta(w)t)}\notag\\&\overset{(a)}{=}2\gamma(w)\lim_{t\rightarrow\infty}\frac{\exp\left(-\left(\frac{\varepsilon^2}{ t}+k_\mathrm{d}t\right)\right)}{-\zeta(w)}\left(\frac{\varepsilon}{\sqrt{t^3}}-\gamma(w)\sqrt{\frac{D_\sigma}{t}}\right)\notag\\&=0,
\end{align}
where step $(a)$ is obtained by applying L'H{\^o}pital's rule. Therefore, we have $\psi_1(\infty,w)=0$ for any value of $\zeta(w)$. According to \eqref{e2}, we obtain $\psi_2(\infty,w)=-2\gamma(w)^2\sqrt{{D_\sigma}/{k_\mathrm{d}}}\exp\left(-\beta\right)$. By substituting $\psi_1(\infty,w)$ and $\psi_2(\infty,w)$ into \eqref{d2}, we obtain $\alpha_2(\infty,w)$. Then, by substituting $\alpha_1(\infty)$ and $\alpha_2(\infty,w)$ into \eqref{H}, we obtain \eqref{Ha}. 

\section{Proof of Theorem \ref{t2}}\label{A3}
As $f_\mathrm{r}(t)$ is the molecule release rate from the MF-based TX membrane at time $t$ and $h_\mathrm{s}(t)$ is the expected molecule hitting rate at the AP-based RX at time $t$ when molecules are uniformly released from the TX membrane at $t=0$, $h_{\ss\mathrm{MF}}(t)$ is obtained as 
\begin{align}\label{hmft}
h_{\ss\mathrm{MF}}(t)=\int_{0}^{t}f_\mathrm{r}(u)h_\mathrm{s}(t-u)\mathrm{d}u,
\end{align}
where $f_\mathrm{r}(t)$ is given by \cite[Eq. (5)]{huang2021fusion}
\begin{align}\label{ff}
f_\mathrm{r}(t)\!=\!\!\!\sum_{n=1}^{\infty}\!\frac{4r_{\ss\T}^2k_\mathrm{f}\lambda_n^3}{2\lambda_nr_{\ss\T}\!-\!\mathrm{sin}\left(2\lambda_nr_{\ss\T}\right)}j_0(\lambda_nr_{\ss\T})\!\exp\!\left(-D_\mathrm{v}\lambda_n^2t\right).
\end{align}

By substituting \eqref{ff} and \eqref{hs} into \eqref{hmft}, we obtain \eqref{hmm}.

\section{Proof of Corollary \ref{c7}}\label{A2}
We note that $H_{\ss\mathrm{MF}}(t)$ can be represented as $H_{\ss\mathrm{MF}}(t)=f_\mathrm{r}(t)*H_\mathrm{s}(t)$. By performing the Laplace transform of both sides, we obtain $\mathcal{H}_{\ss\mathrm{MF}}(s)=\hat{f}_\mathrm{r}(s)\mathcal{H}_\mathrm{s}(s)$, where $\mathcal{H}_{\ss\mathrm{MF}}(s)$, $\hat{f}_\mathrm{r}(s)$, and $\mathcal{H}_\mathrm{s}(s)$ are the Laplace transforms of $H_{\ss\mathrm{MF}}(t)$, $f_\mathrm{r}(t)$, and $H_\mathrm{s}(t)$, respectively. According to \cite[Eqs. (1.8), (5.1), (17.68)]{oberhettinger2012tables}, we obtain
\begin{align}\label{hf}
\hat{f}_\mathrm{r}(s)=\sum_{n=1}^{\infty}\frac{4r_{\ss\T}^2k_\mathrm{f}\lambda_n^3j_0(\lambda_nr_{\ss\T})}{\left(2\lambda_nr_{\ss\T}-\mathrm{sin}\left(2\lambda_nr_{\ss\T}\right)\right)\left(s+D_\mathrm{v}\lambda_n^2\right)},
\end{align}
and 
\begin{align}\label{Hs}
	\mathcal{H}_\mathrm{s}(s)\!=\!\frac{r_{\ss\R}w_\mathrm{e}}{2r_{\ss\T}r_0\zeta(w_\mathrm{e})}\!\!\left[\hat{\xi}_2(s,r_0\!-\!r_{\ss\T}-\!r_{\ss\R})\!-\hat{\xi}_2(s, r_0\!+\!r_{\ss\T}\!-\!r_{\ss\R}\!)\!\right],
\end{align}
where $\hat{\xi}_2(s,z)$ is given in \eqref{hx} on the top of next page.
\begin{figure*}[!t]
	\begin{align}\label{hx}
	&\hat{\xi}_2(s,z)=\exp\left(\gamma(w_\mathrm{e})z\right)\frac{\exp(-\frac{z}{2D_\sigma}(\varpi(w_\mathrm{e})+\sqrt{s+k_\mathrm{d}}))}{\sqrt{s+k_\mathrm{d}}(\varpi(w_\mathrm{e})+\sqrt{s+k_\mathrm{d}})}-\frac{1}{2\sqrt{k_\mathrm{d}}}\exp\left(-z\sqrt{\frac{k_\mathrm{d}}{D_\sigma}}\right)\left[\left(\varpi(w_\mathrm{e})-\sqrt{k_\mathrm{d}}\right)\left(\frac{1}{s}\right.\right.\notag\\&\left.\left.-\frac{\exp(-\frac{z}{D_\sigma}(\sqrt{s+k_\mathrm{d}}\!-\!\sqrt{k_\mathrm{d}}))}{\sqrt{s\!+\!k_\mathrm{d}}(\sqrt{s+k_\mathrm{d}}\!-\!\sqrt{k_\mathrm{d}})}\right)\!-\!\left(\!\varpi(w_\mathrm{e})\!+\!\sqrt{k_\mathrm{d}}\right)\!\!\left(\!\!1\!-\!\exp\!\left(\!2z\sqrt{\frac{k_\mathrm{d}}{D_\sigma}}\right)\!\frac{\exp(-\frac{z}{\sqrt{D}}(\sqrt{k_\mathrm{d}+s}+\sqrt{k_\mathrm{d}}))}{(\sqrt{s+k_\mathrm{d}}+\sqrt{k_\mathrm{d}})\sqrt{s+k_\mathrm{d}}}\right)\right]-\frac{\exp\left(-z\sqrt{\frac{k_\mathrm{d}}{D_\sigma}}\right)}{s}.
	\end{align}
	\hrulefill \vspace*{-1pt}
\end{figure*}
Based on the final value theorem \cite[Eq. (1)]{chen2007final}, we have $\lim\limits_{t\rightarrow\infty}H_{\ss\mathrm{MF}}(t)=\lim\limits_{s\rightarrow 0}s\mathcal{H}_{\ss\mathrm{MF}}(s)$. By substituting \eqref{hf} and \eqref{Hs} into this equation, we obtain \eqref{hm}.

\bibliographystyle{IEEEtran}
\bibliography{ref2}
\end{document}